\documentclass[journal]{IEEEtran}

\usepackage{hhline}
\usepackage[numbers]{natbib}
\usepackage[utf8]{inputenc}
\usepackage{siunitx}
\usepackage{graphicx}
\usepackage[ruled,vlined]{algorithm2e}
\usepackage{makecell}
\usepackage{subfig}
\usepackage{comment}
\usepackage[dvipsnames]{xcolor}

\begin{document}
\title{CyberSpec: Intelligent Behavioral Fingerprinting to Detect Attacks on Crowdsensing Spectrum Sensors}

\author{Alberto Huertas Celdr\'an$^{*1}$, Pedro Miguel S\'anchez S\'anchez$^{2}$, G\'er\^ome Bovet$^{3}$, \\Gregorio Mart\'inez P\'erez$^{2}$, and Burkhard Stiller$^{1}$

\thanks{$^{*}$Corresponding author.}

\thanks{$^{1}$Alberto Huertas Celdr\'an and Burkhard Stiller are with the Communication Systems Group (CSG) at the Department of Informatics (IfI), University of Zurich UZH, 8050 Zürich, Switzerland {\tt\small (e-mail: huertas@ifi.uzh.ch; stiller@ifi.uzh.ch}).}
\thanks{$^{2}$Pedro Miguel S\'anchez S\'anchez and Gregorio Mart\'inez P\'erez are with the Department of Information and Communications Engineering, University of Murcia, 30100 Murcia, Spain {\tt\small (pedromiguel.sanchez@um.es; gregorio@um.es)}.}%
\thanks{$^{3}$G\'{e}r\^{o}me Bovet is with the Cyber-Defence Campus within armasuisse Science \& Technology, 3602 Thun, Switzerland {\tt\small (gerome.bovet@armasuisse.ch)}.}}


\markboth{IEEE Transactions on Dependable and Secure Computing}%
{Shell \MakeLowercase{\textit{et al.}}: Bare Demo of IEEEtran.cls for IEEE Journals}

\maketitle

\begin{abstract}
Integrated sensing and communication (ISAC) is a novel paradigm using crowdsensing spectrum sensors to help with the management of spectrum scarcity. However, well-known vulnerabilities of resource-constrained spectrum sensors and the possibility of being manipulated by users with physical access complicate their protection against spectrum sensing data falsification (SSDF) attacks. Most recent literature suggests using behavioral fingerprinting and Machine/Deep Learning (ML/DL) for improving similar cybersecurity issues. Nevertheless, the applicability of these techniques in resource-constrained devices, the impact of attacks affecting spectrum data integrity, and the performance and scalability of models suitable for heterogeneous sensors types are still open challenges. To improve limitations, this work presents seven SSDF attacks affecting spectrum sensors and introduces CyberSpec, an ML/DL-oriented framework using device behavioral fingerprinting to detect anomalies produced by SSDF attacks affecting resource-constrained spectrum sensors. CyberSpec has been implemented and validated in ElectroSense, a real crowdsensing RF monitoring platform where several configurations of the proposed SSDF attacks have been executed in different sensors. A pool of experiments with different unsupervised ML/DL-based models has demonstrated the suitability of CyberSpec detecting the previous attacks within an acceptable timeframe.
\end{abstract}


\IEEEpeerreviewmaketitle

\section{Introduction}
\label{sec:intro}

%Evolution of IoT and crowdsensing-crowdsourcing
%\IEEEPARstart{T}{he} evolution of mobile communications and the Internet-of-Things (IoT) paradigm is dramatically increasing the number of resource-constrained devices wirelessly connected to the internet. Nowadays, there are about 14 billion IoT devices, and previsions estimate an increment until 64 billion by 2025~\cite{riad:2020:dynamic}. This evolution has brought the crowdsensing paradigm to reality, offering a rich catalog of platforms acquiring data sensed by a relatively large and open number of IoT devices acting as sensors, known as crowd, to later analyze and provide valuable information and/or services. Real examples of crowdsensing platforms can be found in different application scenarios such as car navigation with real-time traffic, aircraft traffic information, or radio frequency (RF) spectrum monitoring, among others~\cite{tang:crowd:2018}.

%Importance of RF Monitors
\IEEEPARstart{T}{he} growth of wireless IoT devices is increasing the demand for radio frequency (RF) spectrum~\cite{riad:2020:dynamic}. Nowadays, while some RF bands are congested (as the ISM band, used by Wi-Fi, Bluetooth, and other standards), others are under-occupied. Consequently, interference occurring in overcrowded RF bands degrade service quality due to the necessity of re-transmissions and the reduction of effective data rates. This situation promotes crowdsensing spectrum sensors as one of the most attractive and valuable solution for integrated sensing and communication (ISAC). Crowdsensing RF monitoring platforms, like ElectroSense~\cite{rajendran:2018:electrosense}, sense the RF spectrum occupancy and allow Cognitive Radio Networks (CRNs) to assign under-occupied bands to particular devices, balancing and optimizing the RF spectrum usage. Furthermore, these platforms are also critical when \textit{(i)} intercepting illegal communications occurring in licensed bands; \textit{(ii)} detecting cyberattacks, jamming or interfering critical transmissions; or \textit{(iii)} classifying transmission standards and technologies.

%Vulnerabilities of resource-constrained devices used in these platforms: External an insider attacks
All these previous tasks highly rely on sensed RF spectrum data, a critical asset that must be protected to ensure trusted services. This protection is a challenging task, since Primary User Emulation Attack (PUEA) and Spectrum Sensing Data Falsification Attack (SSDF) are two of the most well-known and impactful attack families affecting spectrum sensing~\cite{li:2015:survey}. Moreover, this situation becomes even more challenging in the crowdsensing paradigm, where resource-constrained spectrum sensors show well-known and exploitable vulnerabilities, apart from being easily manipulated by users with physical access to them~\cite{meneghello:2019:threats}. 

%Behavioral fingerprinting and AI as a potential solution
Behavioral fingerprinting turned out to be one of the most promising approaches to detect cyberattacks~\cite{bezawada:behavioral:2018}. In such a context, heterogeneous data sources, like hardware events, system calls, or resource usage, can be monitored to create precise device behavioral fingerprints~\cite{sanchez:2020:survey}. These fingerprints are evaluated to detect anomalies by different techniques, being Machine and Deep Learning (ML and DL) the most promising approaches nowadays. However, the following main open challenges appear when combining fingerprinting and ML/DL techniques with spectrum sensing: \textit{(i)} there is no formal definition of the behavior and impact of SSDF attacks affecting the integrity of spectrum data, \textit{(ii)} there is no solution measuring the performance and suitability of intelligent behavioral fingerprinting in resource-constrained spectrum sensors, \textit{(iii)} data sources and events precisely modeling the normal and under attack behavior of spectrum sensors have not been studied and specified, and \textit{(iv)} there is no analysis of the detection performance and scalability of ML and DL-based solutions considering the behavior of groups of similar and different spectrum sensors.

%Contributions
To improve these challenges, the main contributions of this work include:

\begin{itemize}
    \item The definition and deployment of seven novel SSDF attacks (Repeat, Mimic, Confusion, Noise, Spoof, Freeze, and Delay) affecting spectrum sensors. These attacks are categorized into three main families focused on: \textit{(i)} simulating non-existing communications, \textit{(ii)} hiding cyberattacks or illegal transmissions, and \textit{(iii)} performing both actions in parallel.
    
    \item The design and complete implementation of CyberSpec, an ML/DL-oriented framework that uses intelligent behavioral fingerprinting to detect anomalies in resource-constrained spectrum sensors affected by SSDF attacks. Events belonging to the device virtual memory, file system, CPU, network, scheduler, device drivers and random numbers generation are monitored periodically to create behavior fingerprints.
    
    %CyberSpec monitors periodically a set of events belonging to IoT spectrum sensors to create device behavior fingerprints. %After that, it uses unsupervised techniques based on ML/DL to detect behavioral anomalies produced by these attacks.
    
    \item The deployment of CyberSpec in a real IoT-based crowdsensing RF monitoring platform called ElectroSense~\cite{rajendran:2018:electrosense}. Nine Raspberry Pis 3 and 4, acting as sensors, were infected with different configurations of the seven SSDF attacks to later measure the CyberSpec detection performance and time. 
    
    \item The analysis of a pool of experiments to detect normal and under-attack behaviors using ML/DL models per \textit{(i)} individual sensors, \textit{(ii)} families of sensors excluding a different number of devices from training, and \textit{(ii)} combinations of families. The obtained results show that five of the seven SSDF attacks are almost perfectly detected in the three experiments. Device-type and global ML/DL models provide similar detection performance to individual models. Finally, when the 15\%, 33\%, and 50\% of devices are excluded from training, device-type models perform acceptably for most excluded devices (80-70\% TPR and 100\% TNR for five attacks).
\end{itemize}

%Paper structure
The remainder of this article is organized as follows. Section~\ref{sec:related} reviews attacks affecting spectrum data and behavior fingerprinting to detect different cyberattacks. While Section~\ref{sec:scenario} introduces the basics of spectrum sensing, Section~\ref{sec:attacks} presents seven SSDF attacks affecting sensors. Section~\ref{sec:framework} introduces CyberSpec, an ML/DL-oriented framework detecting anomalies. Section~\ref{sec:validation} outlines the deployment and performance of CyberSpec in ElectroSense. Finally, Section~\ref{sec:conclusions} draws conclusions and next steps.

\section{Related Work}
\label{sec:related}

Based on the review of existing cyberattacks affecting spectrum sensing, solutions using device behavioral fingerprinting to detect cyberattacks are analyzed.

\subsection{Cyberattacks Affecting Spectrum Sensing}

Spectrum sensing is recognized as the basic functionality of Cognitive Radio Networks (CRNs), and it is vulnerable to different cyberattacks. In~\cite{alrabaee:2014:attacks}, a series of malicious behaviors are categorized into \textit{(i)} misbehaving, \textit{(ii)} selfish, \textit{(iii)} cheating, or \textit{(iv)} malicious. Misbehaving consists of breaking the rules established by the network or spectrum sensing platform. Selfish behavior is to keep the network resources for the attacker's benefit. A cheating behavior occurs when the attacker provides fake spectrum information to increase his/her quality of service (QoS). Finally, malicious behavior is when the attacker targets the spectrum sensing to degrade the other nodes QoS and the network efficiency~\cite{khasawneh:2017:attCRN}.

Some of these malicious behaviors can be shown by different cyberattacks affecting the sensing phase of spectrum monitors. This is the case of Primary User Emulation Attack (PUEA) and Spectrum Sensing Data Falsification (SSDF)~\cite{khasawneh:2017:attCRN}. PUEA~\cite{vivekanand:2020:puea} simulate transmissions of primary users (PU) over licensed spectrum bands. On the other hand, SSDF attacks~\cite{yadav:2020:ssdf} send fake sensing data to the sensing platform and combine misbehaving, cheating, and selfish behaviors. Apart from these two attacks, others can be launched as a result of PUEA and/or SSDF~\cite{li:2015:survey}. Denial-of-Service (DoS) attacks are one of them, where the malicious spectrum sensor emulates a transmission from a PU to force secondary users (SU) to vacate the spectrum. The DoS attack results in degrading the QoS of SU~\cite{vinko:2014:dos}.

%Conclusion
Despite the progress done in the literature regarding SSDF attacks affecting RF monitoring platforms, there is a lack of work detailing particular behaviors of different SSDF attacks and their impacts on the spectrum data.

\subsection{Cyberattack Detection Through Behavioral Fingerprinting}

% IoT device fingerprinting

The number of existing works applying behavioral fingerprinting in IoT devices from the host perspective (as this work does) is reduced, but one of them is HADES-IoT~\cite{breitenbacher:2019:hades}. HADES-IoT is a host-based anomaly detection system for different Linux-based IoT devices that creates white lists of benign system calls. A 100\% effectiveness is reported when evaluating the system with different IoT malware affecting seven different IoT devices. Also, in the field of fingerprinting and IoT, the authors of~\cite{ngo:2020:adaptive} propose a layered and adaptive AD solution for IoT that uses Hierarchical Edge Computing (HEC) and three DL models with different levels of complexity. IoT devices, edge servers, and cloud architectures are the layers of the HEC where the models are deployed and evaluated depending on contextual information. A pool of experiments demonstrates a reduced detection delay while maintaining acceptable detection accuracy. Finally, DAIMD~\cite{jeon:2020:hybridIoT} defines a hybrid scheme that monitors memory, network, virtual file system, process, and system calls of devices to detect both well-known and zero-day attacks. A convolution neural network (CNN) model is trained to classify samples into benign and malicious ones successfully.

% General Device fingerprinting -> Syscalls
System calls, resource usage, or Hardware Performance Counters (HPC) are widely used to create behavioral fingerprints and detect cyberattacks affecting different devices. From the system calls perspective, VizMal~\cite{delorenzo:2020:vizmal} detects Android malware by creating color boxes that represent software execution time windows. The color of each box refers to the maliciousness level, and the size is the number of system calls executed during the time window. A Long Short-Term Memory (LSTM) Neural Network trained with samples labeled as malware and non-malware provides promising detection results. VMGuard~\cite{mishra:2020:vmguard} is another ML-oriented security system that uses system calls to detect malware affecting Virtual Machines (VM) in cloud scenarios. VMGuard monitors the system calls of VMs to create a `Bag of n-grams (BonG)' integrated with the Term Frequency-Inverse Document Frequency (TF-IDF) method. During the evaluation process, Random Forest classifies different attacks successfully.

\begin{table*}
	\caption{Comparison of Solutions Using Behavioral Fingerprinting to Detect Cyberattacks}
	\begin{tabular}{p{2cm}|p{0.6cm}|p{2.6cm}|p{3.3cm}|p{2.4cm}|p{2.2cm}|p{2cm}}
		\textbf{Solution} & \textbf{Year} & \textbf{Scenario} & \textbf{Data Source} & \textbf{Attack} & \textbf{Technique} & \textbf{Performance} \\ \hline
		HADES-IoT~\cite{breitenbacher:2019:hades} & 2019 & IoT Devices (Linux) & Syscalls & IoT Malware & White List & 1 Accuracy \\ \hline
		DAIMD~\cite{jeon:2020:hybridIoT} & 2020 & IoT Devices & Memory, Network, File System, Processes, and Syscalls & IoT Malware & CNN & 0.998 Accuracy\\ \hline
		VizMal~\cite{delorenzo:2020:vizmal} & 2020 & Smartphone (Android) & Syscalls & Android Malware & LSTM & 0.098 FPR\\ \hline
		VMGuard~\cite{mishra:2020:vmguard} & 2020 & Cloud Architectures (VMs) & Syscalls & Malware (UNM Dataset) & Random Forest & 0.93 - 0.99 Accuracy\\ \hline
		Ravichandiran et al.~\cite{ravichandiran:2018:scale} & 2018 & Cloud Architecture (Micro-services) & Resource Usage (CPU) & DoS & Statistical Models & 1 Accuracy\\ \hline
		RADS \cite{barbhuiya:2018:rads} & 2018 & Cloud Data Center (VMs)  & Resource Usage (CPU) & DDoS and Cryptominers & One Class Classification & 0.90 - 0.95 Accuracy\\ \hline
		Karl et al.~\cite{ott:2018:hpc} & 2018 & Embedded Systems & HPCs & Malicious Intrusions & LSTM and HMM & 0.98 - 1 Accuracy\\ \hline
        CyberSpec (this work) & 2021 & IoT Spectrum Sensors (Raspberry Pi) & Linux kernel events %HPCs and Resource Usage 
        & SSDF Attacks & Autoencoder & 0.60 - 1 TPR, 0.90 - 0.98 TNR \\ \hline
	\end{tabular}
	\label{tab:comparison}
\end{table*}

% General Device fingerprinting -> Resource Usage & HPC
Resource usage is another well-known and widely used data source to create behavioral fingerprints. In this context, the solution presented in~\cite{ravichandiran:2018:scale} detects Denial-of-Service (DoS) attacks in cloud scenarios by considering CPU usage. CPU usage statistics of micro-services running on cloud architectures are monitored to later detect DoS attacks using auto-regressive statistical models. RADS~\cite{barbhuiya:2018:rads} is another system that monitors resource usage to detect DDoS attacks in cloud data centers. RADS uses one class classification algorithm and time series analysis to obtain 90-95\% accuracy with a low false-positive rate of 0-3\% when detecting anomalies produced by DDoS and cryptomining attacks. Finally, HPCs are considered in \cite{ott:2018:hpc}, where authors propose an ML/DL-oriented solution that detects deviations on normal behaviors produced by malware affecting embedded systems with a repetitive functionality. The validation reports an accuracy higher than 95\% with Hidden Markov Models (HMM) and LSTM Neural Networks.

%Conclusion
\tablename~\ref{tab:comparison} compares the most representative aspects of related work. In conclusion of this analysis, most of the solutions are designed for powerful devices, being useless for resource-constrained ones due to the monitored data sources, used techniques and attacks. Despite HADES-IoT and DAIMD work for IoT, they do not evaluate their suitability or performance when detecting SSDF attacks in spectrum sensors (having a particular functionality that affects the device behavior). Finally, none of the analyzed works studied the detection performance of ML/DL models combining behaviors of different types of devices. Due to the previous facts, more work aligned with SSDF attacks affecting RF sensing platforms and ML/DL-based detection frameworks is needed. %To finalize, it is worth noting that due to major differences between existing work and CyberSpec in terms of devices, attacks, and application scenarios, it is not possible to compare their detection performance.

%------------------------------------------------------

\section{Crowdsensing Spectrum Sensors \& Cybersecurity Issues}
\label{sec:scenario}

Crowdsensing spectrum sensors are resource-constrained devices, such as a Raspberry Pi, equipped with Software-defined Radio (SDR) kits. Typically, spectrum sensors implement a process in charge of \textit{(i)} dividing the spectrum into fixed-size segments, and \textit{(ii)} scanning cyclically the segments composing the whole RF spectrum, from the lower frequency supported by the sensor to the higher. This RF monitoring process enables the collection of different data types per segment, such as Power Spectrum Data (PSD), indicating the spectrum occupancy of each particular segment. After that, the collected data is sent periodically to a central platform, in charge of processing and analyzing it to provide services in charge of \textit{(i)} optimizing RF occupancy, \textit{(i)} intercepting illegal communications over licensed bands, and \textit{(iii)} detecting cyberattacks jamming, interfering legit communications.

The trustworthiness of the previous services depends on the integrity of PSD collected by IoT spectrum sensors. However, the nature of crowdsensing RF monitoring platforms, relying on resource-constrained spectrum sensors vulnerable to cyberattacks and manipulations of dishonest users, makes spectrum sensors vulnerable to SSDF attacks. In such a context, and as explained in Section~\ref{sec:related}, there is a literature gap in terms of defining and implementing SSDF attacks having different behaviors and impacts on IoT spectrum sensors.

\section{SSDF Attacks}
\label{sec:attacks}

\begin{figure}[ht]
    \centering
    \includegraphics[width=0.8\columnwidth]{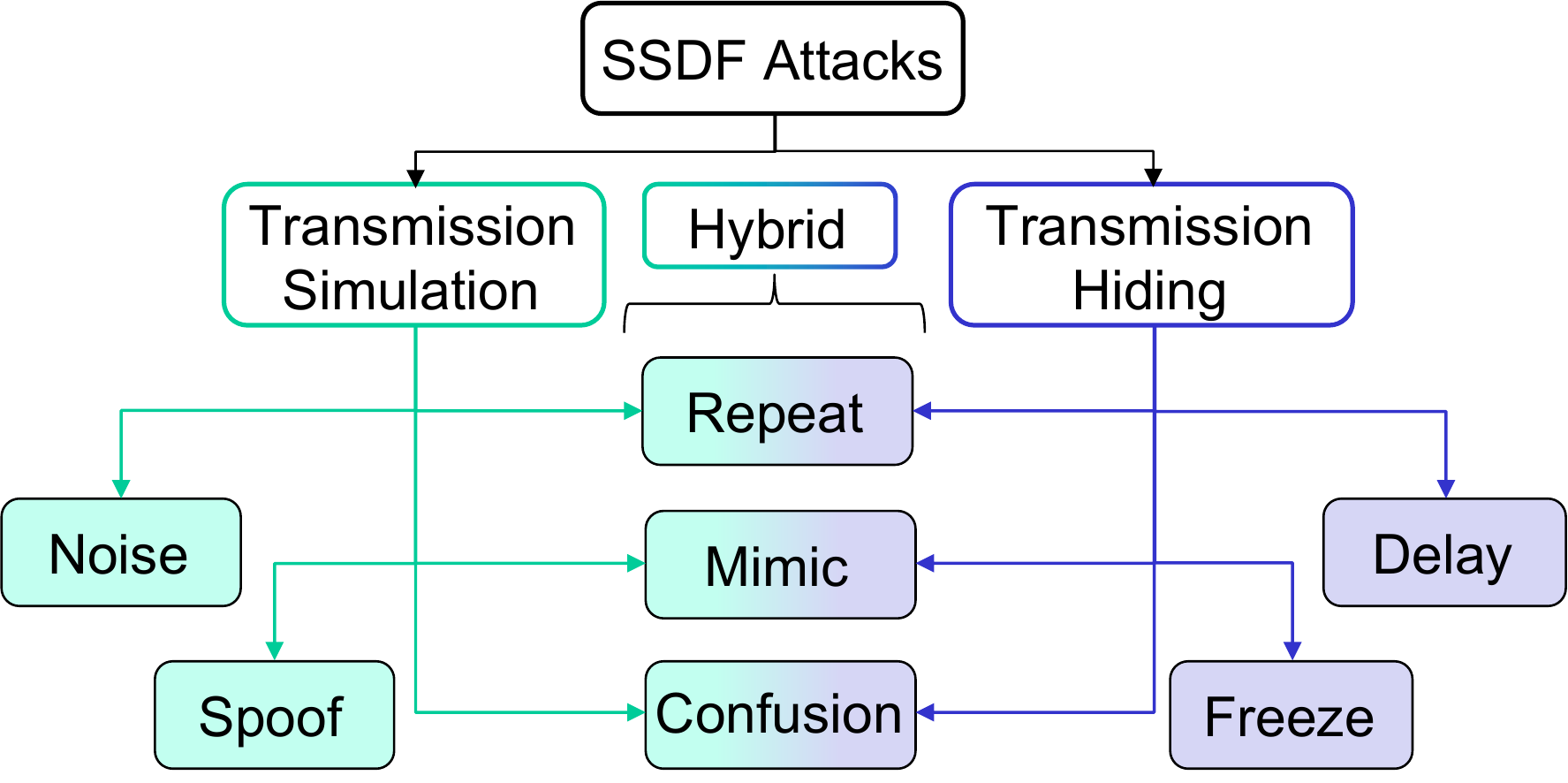}
    \caption{Taxonomy of Proposed SSDF Attacks %Affecting the Integrity of RF Spectrum Data
    }
    \label{fig:attacks}
\end{figure}

Seven novel SSDF attacks showing differences in terms of PSD impact are presented in this section. The proposed SSDF attacks are categorized into three main families: \textit{Transmission Simulation}, \textit{Transmission Hiding}, and \textit{Hybrid}. Transmission Simulation attacks modify the PSD of spectrum segments to simulate non-existing wireless transmissions, since the goal of these attacks is to keep spectrum segments (those where non-existing transmissions are simulated) under-occupied to be used by the attacker. Secondly, they overload the remainder spectrum segments (where real transmissions are assigned) to increase interference and downgrade the Quality-of-Service (QoS). Transmission Hiding attacks are focused on disrupting services that detect cyberattacks or unauthorized transmission by hiding them. Finally, Hybrid attacks can cover the previous two families, simulating fictitious transmissions, hiding illegal ones, or doing both in parallel. \figurename~\ref{fig:attacks} shows the attacks belonging to each family.

\subsection{Hybrid SSDF Attacks}

Attacks belonging to this family can be used either to simulate transmissions or hide them. The family is composed of the following three attacks: \textit{Repeat}, \textit{Mimic}, and \textit{Confusion}.

\begin{figure}[ht]
    \centering
    \includegraphics[width=0.7\columnwidth]{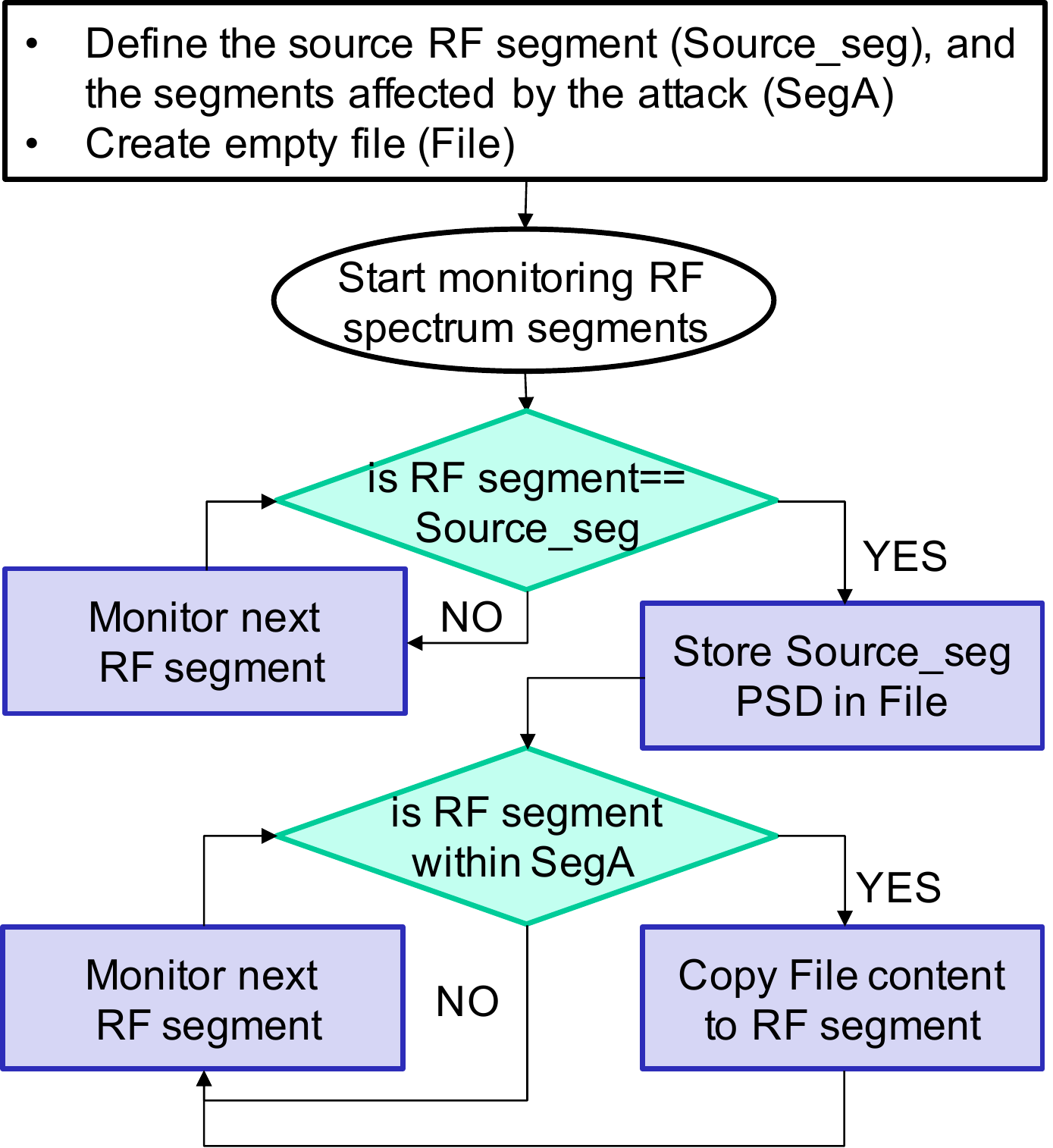}
    \caption{Diagram Flow of Repeat Attacks}
    \label{fig:repeat}
\end{figure}

\textbf{Repeat} attacks copy in a particular moment of time the PSD of a selected spectrum segment and continuously replicates it in the segments targeted by the attack. More in detail and as can be seen in \figurename~\ref{fig:repeat}, first, the attacker selects the RF segment that wants to be replicated (Source\_seg) and the spectrum segments whose PSD are going to be manipulated (SegA). Depending on the PSD values of Source\_seg, the attack will hide or simulate transmissions. After that, it creates a file (File) to save PSD values of Source\_seg. At this point, the RF spectrum is continuously scanned, as indicated in Section~\ref{sec:scenario}. During the first scanning of the RF spectrum, the PSD values of Source\_seg are stored in File (only once). In the next RF scanning cycles, the attack modifies the PSD values of the SegA with the File content. %This attack impacts PSD values and the internal behavior of the sensor, creating more activity in file system events.

%-------------------

\begin{figure}[h!]
    \centering
    \includegraphics[width=0.95\columnwidth]{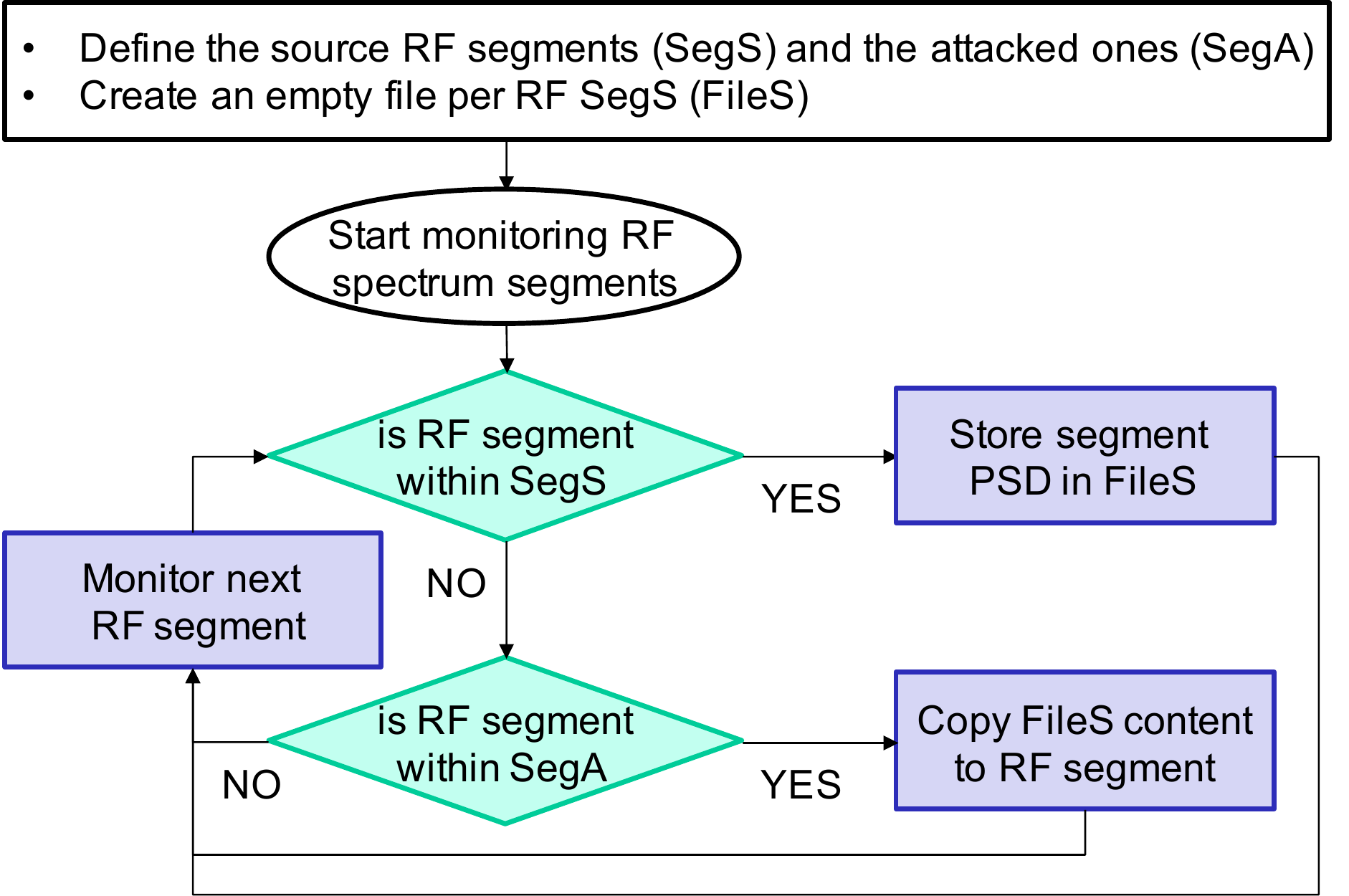}
    \caption{Diagram Flow of Mimic Attacks}
    \label{fig:mimic}
\end{figure}

\textbf{Mimic} attacks are an evolution of Repeat, being the creation of new PSD copies per RF spectrum the main difference of them. In particular, as seen in \figurename~\ref{fig:mimic}, this attack defines two sets of spectrum segments (SegS are the segments whose PSDs are going to be replicated, and SegA the replaced segments) and creates one empty file (FileS) per SegS. After that, the scanning process starts, and for each RF spectrum cycle the attack stores the PSD values of SegS in FileS. Once the files contain the PSD values of SegS, the attack substitutes the PSD of the SegA with the files content. The number of RF segments of SegS and SegA must be equal, and depending on the SegA occupancy the attack simulates or hides transmissions. %Thus, this attack impacts the sensor internal behavior, particularly the number of read and write operations.

%-------------------

\begin{figure}[ht]
    \centering
    \includegraphics[width=0.8\columnwidth]{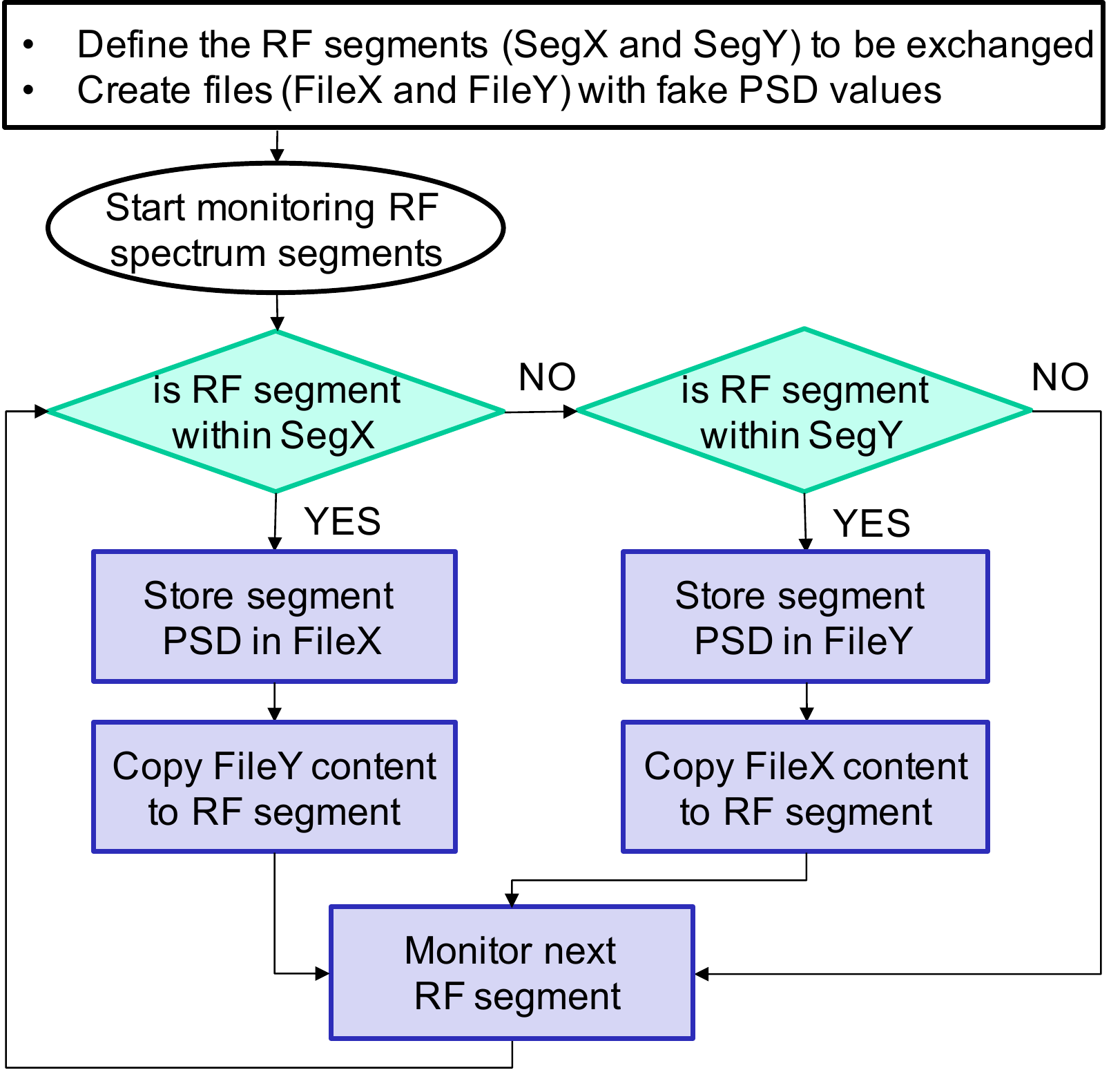}
    \caption{Diagram Flow of Confusion Attacks}
    \label{fig:confusion}
\end{figure}

\textbf{Confusion} attacks pretend to exchange the occupancy levels of two or more RF segments. The attack can be used to hide a transmission, to simulate an non-existing one, or for both at the same time, creating confusion. \figurename~\ref{fig:confusion} shows how the attack starts defining the sets of spectrum segments that are going to be exchanged (SegX and SegY), and creating two files (FileX and FileY) with PSD values. After that, the monitoring process initiates the sequential RF scanning. When the scanning process senses the occupancy of a segment belonging to the SegX, its PSD values are stored in FileX, and replaced with the content of FileY. Similarly, when the monitor senses an RF segment belonging to SegY, it saves its PSD in FileY and replaces the segment occupancy with FileX content. %The attack affects the internal behavior of the sensor increasing the number of read and write operations.

\subsection{Transmission Simulation SSDF Attacks}

Attacks of this family are in charge of modifying the RF spectrum occupancy data to simulate fake transmissions. This family is composed of two attacks: \textit{Noise} and \textit{Spoof}.

\begin{figure}[ht]
    \centering
    \includegraphics[width=0.70\columnwidth]{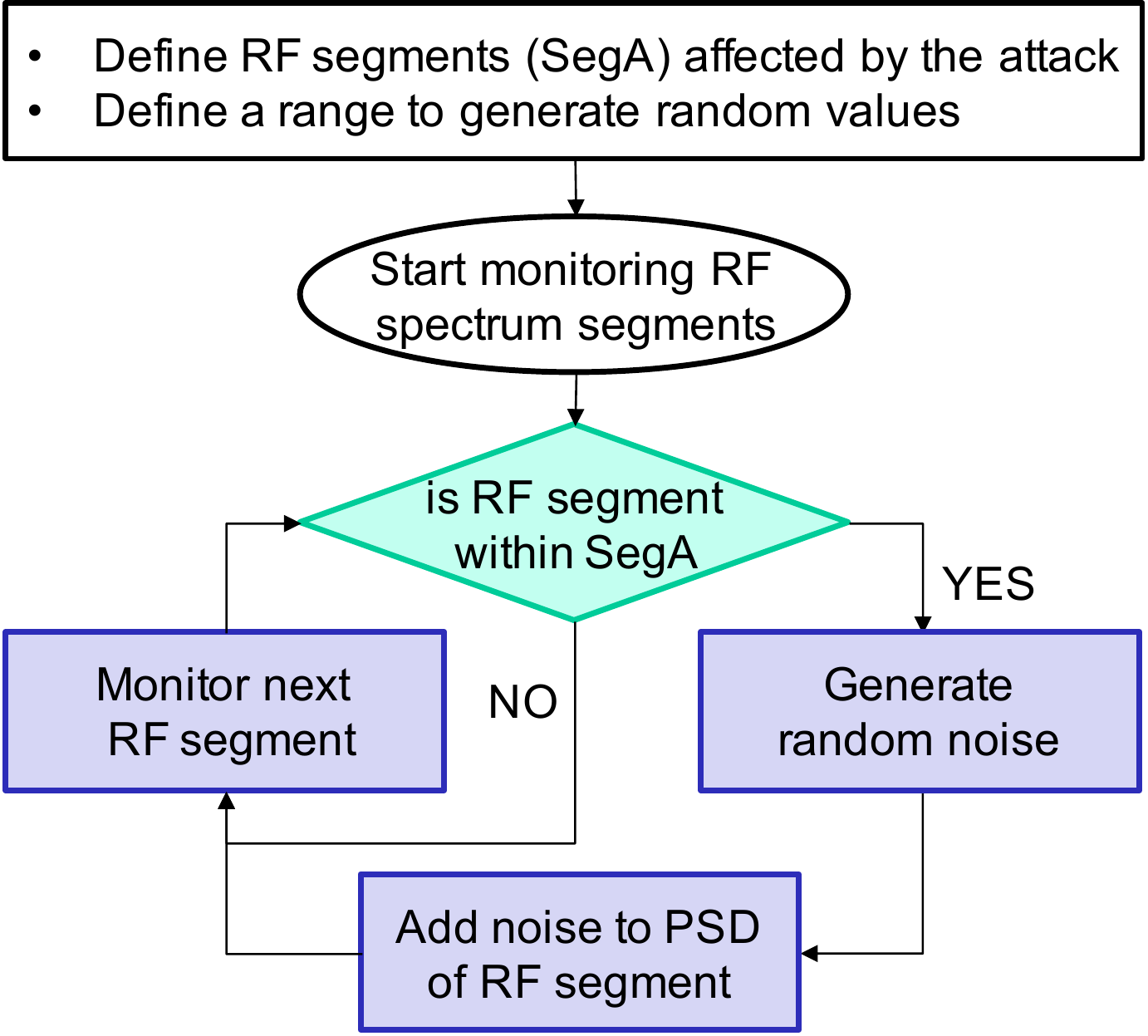}
    \caption{Diagram Flow of Noise Attacks}
    \label{fig:noise}
\end{figure}

\textbf{Noise} attacks focus on adding random noise to the occupancy level of a set of RF spectrum segments affected by the attack. \figurename~\ref{fig:noise} shows the life-cycle of this attack, where the attacker starts defining the spectrum segments (SegA) that will be attacked and the intensity of the noise. After that, the RF monitoring process starts. When those segments affected are scanned, the attack generates random numbers and adds them to their PSD values. This process is periodically repeated in each RF scanning cycle for all segments affected. %Thus, this attack impacts the internal behavior of the IoT spectrum sensor, increasing the generation of random numbers.

%-------------------

%The main goal of spoof attacks consists of adding random noise to the occupancy level of a set of RF segments (ensuring the simulation of transmissions) that are used to replace the occupancy of a set of attacked RF segments.

\begin{figure}[ht]
    \centering
    \includegraphics[width=0.85\columnwidth]{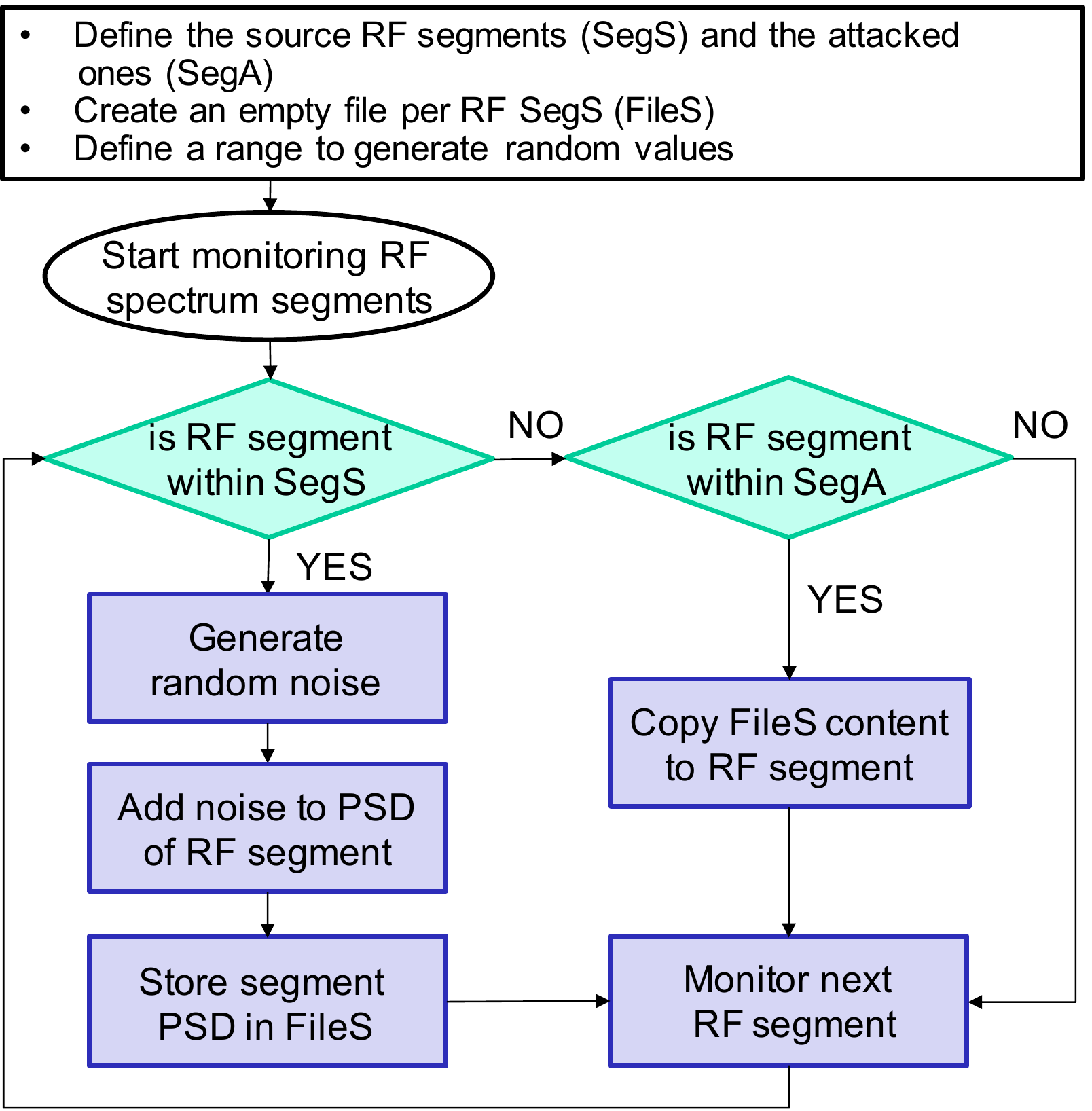}
    \caption{Diagram Flow of Spoof Attacks}
    \label{fig:spoof}
\end{figure}

\textbf{Spoof} attacks are an evolution of mimic attacks. The main difference is that spoof adds random noise to the occupancy level of source spectrum segments to replicate a transmission but adding differences to complicate its detection. As seen in \figurename~\ref{fig:spoof}, the attacker defines the sets of source and attacked spectrum segments (SegS and SegA), creates a file per spectrum segment belonging to SegS (FileS), and defines the intensity of random noise. After that, the RF monitoring process starts. If the current segment belongs to SegS, one random value per PSD is generated, added to the PSD of the current segment, and stored in FileS. When the segment is within the set of the attacked segments (SegA), the occupancy of the current segment is replaced with the content of FileS. This process is repeated for the whole RF spectrum and across time. %This type of attack mainly affects the internal behavior of the IoT spectrum sensor by increasing reads, writes, and random numbers. 

%-------------------

\subsection{Transmission Hiding SSDF Attacks}

\textit{Freeze} and \textit{Delay} are the attacks belonging to this family and pretend to hide illegal or unauthorized transmissions.

\begin{figure}[ht]
    \centering
    \includegraphics[width=0.8\columnwidth]{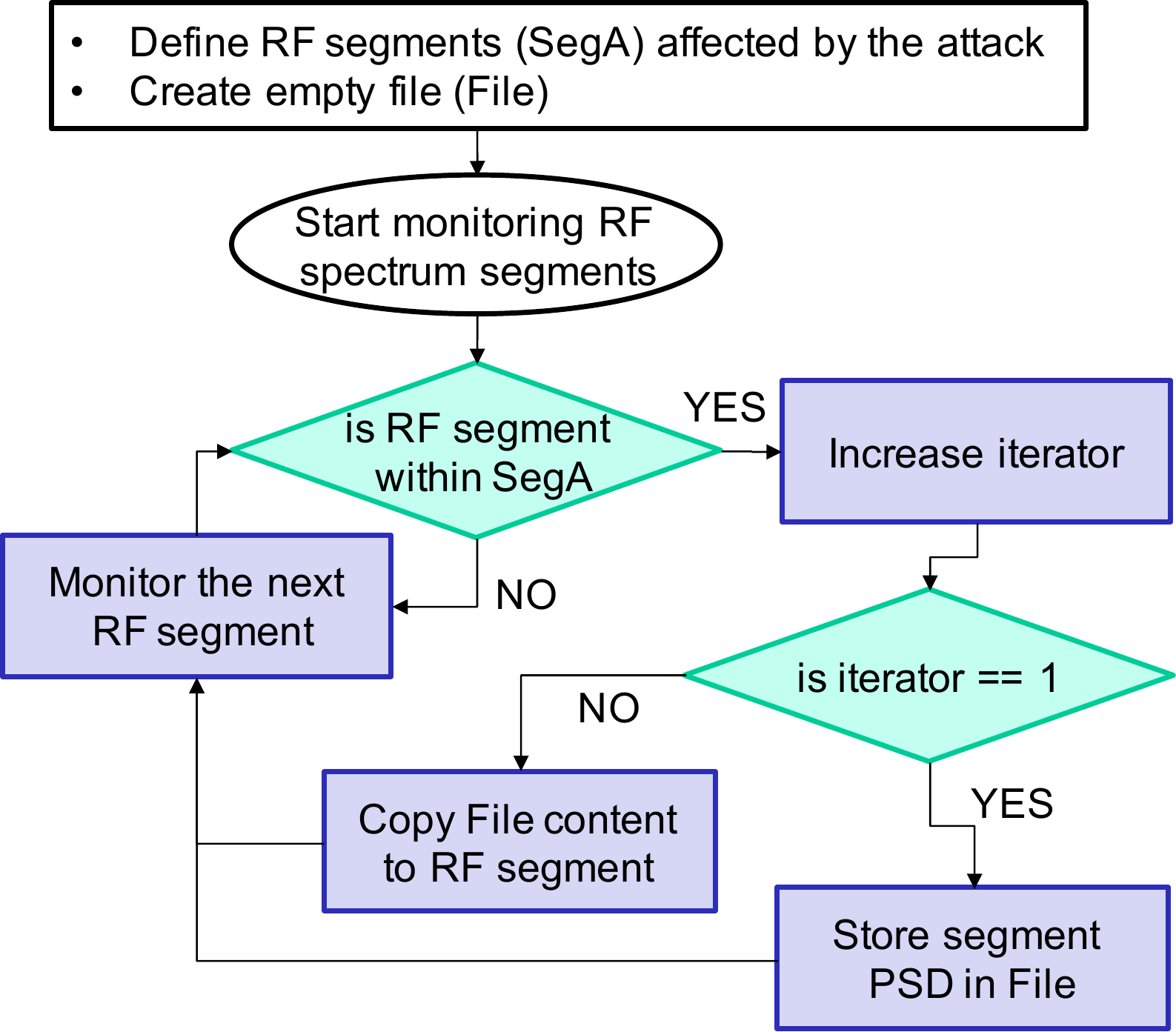}
    \caption{Diagram Flow of Freeze Attacks}
    \label{fig:freeze}
\end{figure}

\textbf{Freeze} attacks copy the PSD values of one or more spectrum segments in a given moment and replicate them across the time. This type of attack hides any transmission starting after the screenshot generation and substitution. As seen in \figurename~\ref{fig:freeze}, the attacker begins defining the spectrum segments affected by the attack (SegA) and creates an empty file, where the PSD values will be stored. After that, the cyclic RF scanning starts. The first time (first scan) the monitor senses the occupancy of a segment belonging to SegA, its PSD values are stored in FileA. In successive iterations of the scanning process, when the monitor senses a segment belonging to SegA, its PSD values will be replaced with the PSDs previously saved in FileA for that particular segment. %This kind of attack impacts the internal behavior of the IoT spectrum sensors, especially the file system, increasing the number of reading operations.

\begin{figure}[h!]
    \centering
    \includegraphics[width=0.75\columnwidth]{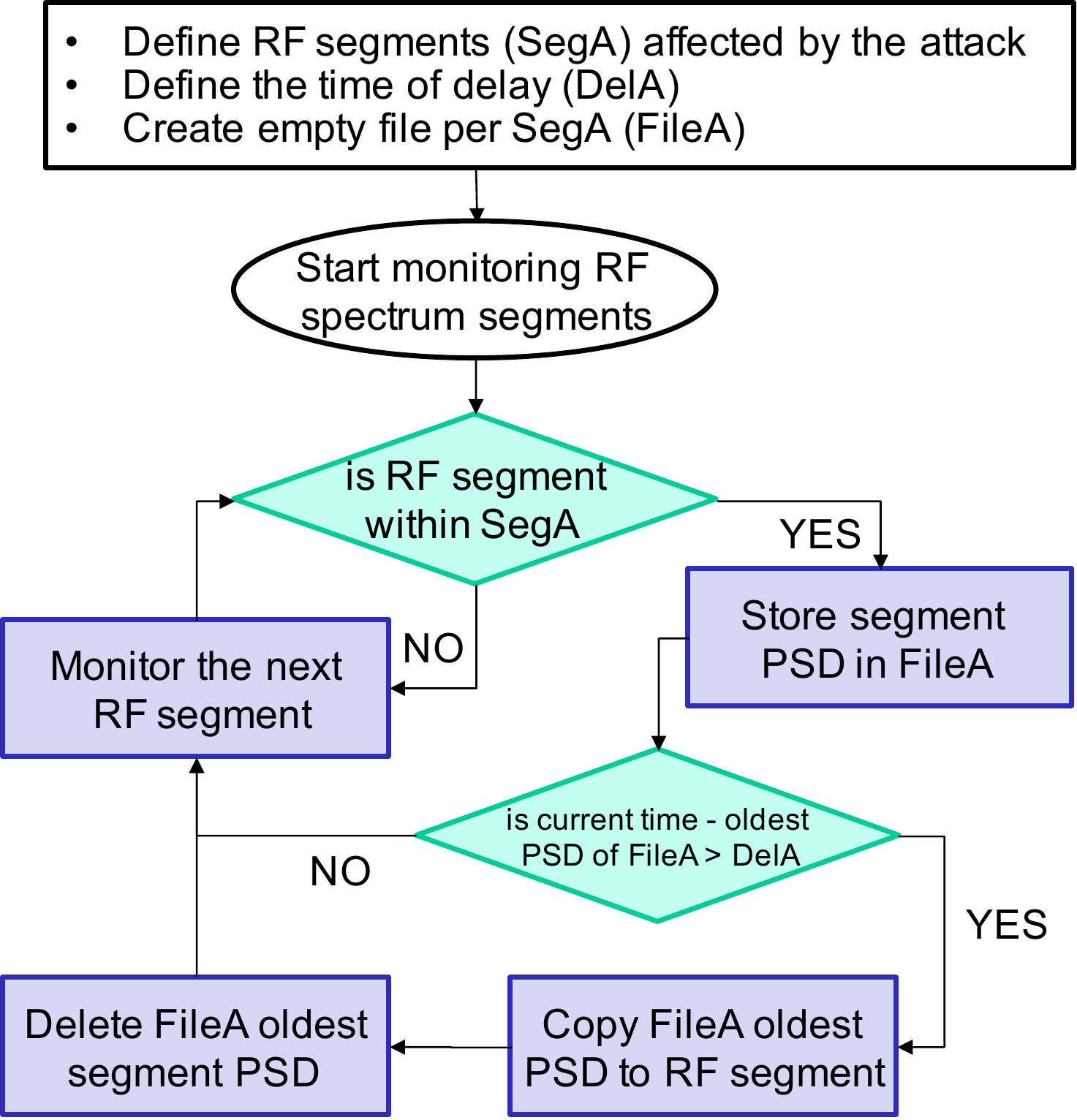}
    \caption{Diagram Flow of Delay Attacks}
    \label{fig:delay}
\end{figure}

\textbf{Delay} attacks are an evolution of Freeze, and their main goal is to send the RF monitoring platform obsolete or non-updated PSDs of affected spectrum segments. The main difference between Delay and Freeze is that Freeze always sends the same occupancy level for each affected segment, while Delay keeps a sliding time window to send different outdated PSD. In this sense, \figurename~\ref{fig:delay} shows that the attack starts defining the set of affected spectrum segments (SegA), the time of delay to provide the segments occupancy (DelA), and creates an empty file per affected segment (FileA). After that, the RF spectrum scanning starts. If the sensed spectrum segments belong to the set of SegA, its PSD values are stored in FileA. Later, it is checked if the time window of delay is over. If so, the segment PSD values are replaced with the oldest content of FileA, and the content is deleted from the file. In contrast, when DelA is not over, the attack stores the segment occupancy values in FileA, but the segment occupancy is not altered. It happens when the attack starts and until the selected delay is reached. %This type of attack is the most elaborated one, and its impact affects different internal sources of IoT spectrum sensors, such as the file system, virtual memory, or system calls.

%--------------------------------------------------------

\section{CyberSpec Framework}
\label{sec:framework}

The CyberSpec framework considers device behavioral fingerprinting to detect anomalies produced by SSDF attacks affecting resource-constrained spectrum monitors. For that, the following two modules compose the flexible framework: 

\begin{itemize}
    \item \textit{Behavior Fingerprinting}. Monitor the internal behavior of the resource-constrained spectrum sensor from different data sources to create behavioral fingerprints.

    \item \textit{Cyberattacks Detection}. Train ML and DL algorithms with the device behavior fingerprints. Moreover, it evaluates the trained models with the real-time behavior of the sensors to detect anomalies produced by cyberattacks.
    %\item \textit{Decision-Making}. Interpret the predictions of the previous module to analyze the cyberattack risks.
    
\end{itemize}

\begin{figure}[ht]
    \centering
    \includegraphics[width=0.8\columnwidth]{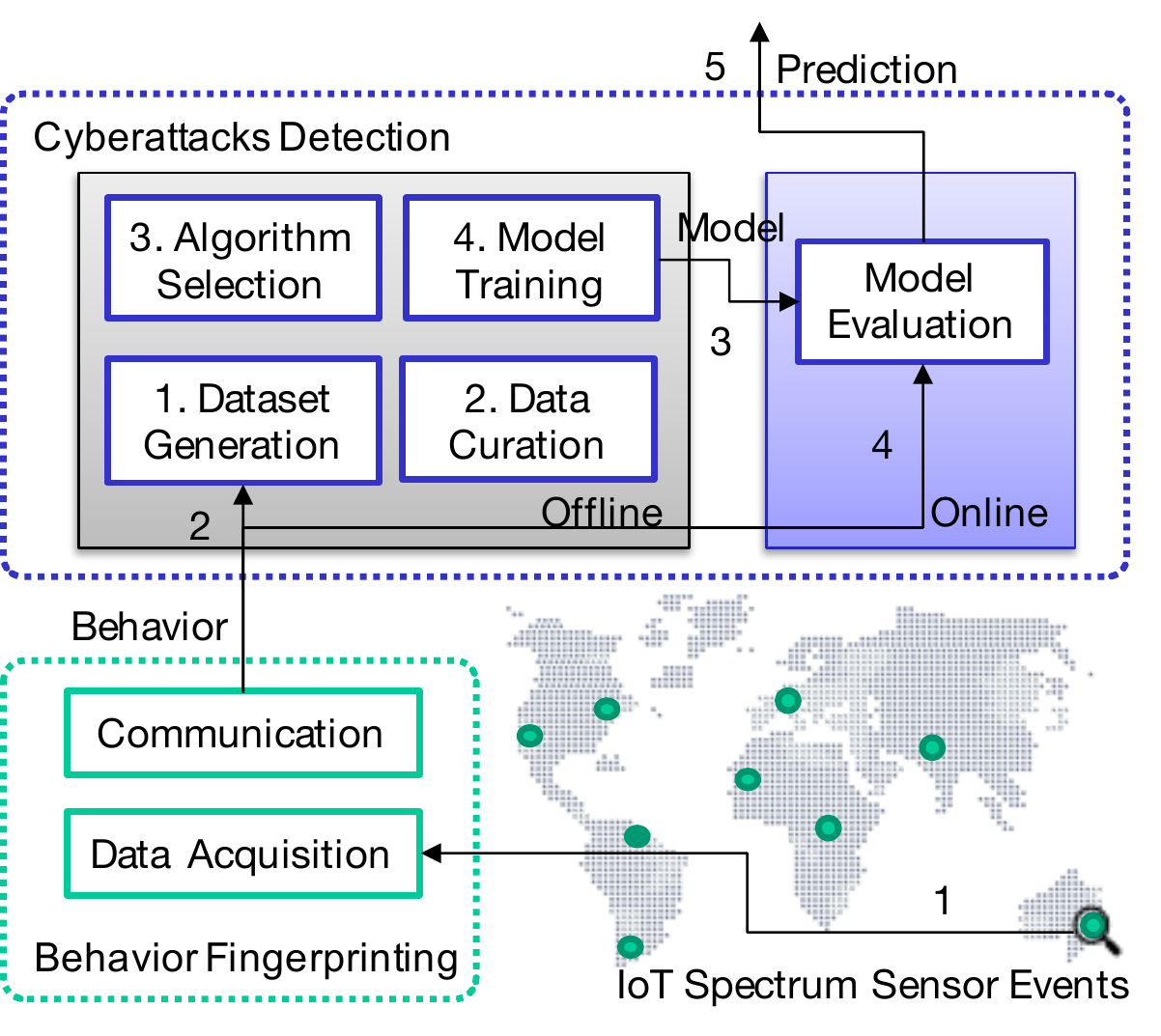}
    \caption{Design of the CyberSpec Framework}
    \label{fig:architecture}
\end{figure}

\figurename~\ref{fig:architecture} shows the two modules of the proposed framework as well as their components, which are explained in this section. CyberSpec is fully deployed in a distributed way, where the spectrum sensor hosts the Behavior Fingerprinting module, and the Cyberattacks Detection is deployed on a server (but it is suitable to be deployed in the sensors).

\subsection{Behavioral Fingerprinting}

This module monitors the behavior of resource-constrained spectrum sensors in periodic and configurable time windows to create behavior fingerprints that are sent to the Cyberattack Detection module. To achieve this functionality, the \textit{Data Acquisition} component monitors all events shown in  \tablename~\ref{tab:features} (step 1 in \figurename~\ref{fig:architecture}). The criteria to select the event families and events relies on covering the most diverse and relevant hardware and software characteristics of resource-constrained spectrum sensors, such as Raspberry Pi models. These events are monitored periodically using configurable time windows (details are provided in Section~\ref{sec:validation}). In particular, the \textit{Perf} Linux command is used for monitoring purposes. Once the events are measured, their values are saved in a behavioral vector containing a timestamp. Finally, the \textit{Communication} component sends the behavioral vector to the Cyberattacks Detection module (step 2 in \figurename~\ref{fig:architecture}). %Algorithm \ref{alg:gathering} shows the pseudo-code implementing the previous functionality.

\begin{comment}
\begin{algorithm}[ht!]
\SetAlgoLined
\KwResult{Send vector with behavior measurements.}
 timestamp = date\;
 monitored\_events = events of \tablename~\ref{tab:features}\;
 time\_window = 50 s\;
 behavior\_vector, events = ""\;
 \While{true}{
 events = \textit{perf stat -e} monitored\_events \textit{-a sleep} time\_window\;
 behavior\_vector = timestamp + events\;
 \textit{curl -X POST -F} behavior\_vector server\_url\;
 }
 \caption{Behavior Fingerprinting Monitoring}
 \label{alg:gathering}
\end{algorithm}
\end{comment}

\subsection{Cyberattacks Detection}

The Cyberattacks Detection module hosts two processes. The first is performed offline and consists of training unsupervised ML/DL models with the normal behavior of the spectrum sensor. Once it is done, an online process periodically evaluates the real-time behavior of the sensors with the trained models to detect anomalies produced by cyberattacks.

The offline process is executed before the online one and it is composed of the following four components: \textit{Dataset Generation}, \textit{Data Curation}, \textit{Algorithm Selection}, and \textit{Model Training}. The Dataset Generation periodically receives data vectors modeling the normal (benign) behavior of each resource-constrained spectrum sensor and creates a dataset for each device. The dataset creation task is configurable, but around ten days are required to build a stable behavioral fingerprint. It is critical to ensure that no attacks affect the device during this time period and contextual noise is minimized. Once the dataset is created, the Data Curation component performs several tasks. The first is to remove noisy vectors, features with constant values, features providing no relevant information, highly correlated features, and features with different data distributions across the time and between sensors. After performing these tasks, the resulting features are shown in \tablename~\ref{tab:features} in blue. Finally, other important tasks of the Data Curation process are i) splitting the datasets into training (72\%), validation (18\%) and testing (10\%), and ii) normalizing the values of the features.  Once the final list of features is decided, the Algorithm Selection component selects several unsupervised ML/DL algorithms (explained in Section~\ref{sec:validation}) to be trained by the framework offline process. After that, the Model Training component receives the selected algorithms and feeds them with the training set to build the models. Finally, the performance of each model is evaluated with the validation set, and the models obtaining the best scores are sent to the online process (\textit{cf.} step 3 in \figurename~\ref{fig:architecture}).

\begin{table*}[ht!]
    \centering
    \scriptsize
    \caption{Features, Categorized into Event Families, Used by CyberSpec to Detect Attacks on IoT Spectrum Sensors. The Final Selected Features are Shown in Blue.}
    \begin{tabular}{m{4em}|m{18em}m{16em}m{18em}}
    \makecell[c]{\textbf{Family}} &  & \textbf{Features} & \\
    \hline
    Network & \makecell[l]{tcp:tcp\_destroy\_sock\\
     \colorbox{blue!15}{tcp:tcp\_probe}\\
     udp:udp\_fail\_queue\_rcv\_skb\\
     \colorbox{blue!15}{net:net\_dev\_queue}\\
     }&
     \makecell[l]{\colorbox{blue!15}{net:net\_dev\_xmit}\\ 
     qdisc:qdisc\_dequeue\\
     skb:consume\_skb\\
     \colorbox{blue!15}{skb:kfree\_skb}\\}&
     \makecell[l]{skb:skb\_copy\_datagram\_iovec\\ sock:inet\_sock\_set\_state\\
     \colorbox{blue!15}{fib:fib\_table\_lookup}\\} \\
    
    \hline
    Virtual Memory & \makecell[l]{\colorbox{blue!15}{vwriteback:global\_dirty\_state}\\  
     \colorbox{blue!15}{writeback:sb\_clear\_inode\_writeback}\\
     \colorbox{blue!15}{writeback:wbc\_writepage}\\
     writeback:writeback\_dirty\_inode\\
     \colorbox{blue!15}{writeback:writeback\_dirty\_inode\_enqueue}\\
     \colorbox{blue!15}{writeback:writeback\_dirty\_page}\\
     \colorbox{blue!15}{writeback:writeback\_mark\_inode\_dirty}\\}&
     \makecell[l]{
     writeback:writeback\_pages\_written\\
     \colorbox{blue!15}{writeback:writeback\_single\_inode}\\
     \colorbox{blue!15}{writeback:writeback\_write\_inode}\\
     \colorbox{blue!15}{writeback:writeback\_written}\\
     kmem:kfree\\
     kmem:kmalloc\\
     \colorbox{blue!15}{kmem:kmem\_cache\_alloc}\\}&
     \makecell[l]{
     \colorbox{blue!15}{kmem:kmem\_cache\_free}\\
     kmem:mm\_page\_alloc\\
     kmem:mm\_page\_alloc\_zone\_locked\\
     \colorbox{blue!15}{kmem:mm\_page\_free}\\
     \colorbox{blue!15}{kmem:mm\_page\_pcpu\_drain}\\
     \colorbox{blue!15}{page-faults}\\
     \colorbox{blue!15}{pagemap:mm\_lru\_insertion}\\}\\

    \hline
    \makecell[l]{File\\Systems} & \makecell[l]{\colorbox{blue!15}{jbd2:jbd2\_handle\_start}\\  
     \colorbox{blue!15}{jbd2:jbd2\_start\_commit}\\
     block:block\_bio\_backmerge\\
     \colorbox{blue!15}{block:block\_bio\_remap}\\}&
     \makecell[l]{
     \colorbox{blue!15}{block:block\_dirty\_buffer}\\
     \colorbox{blue!15}{block:block\_getrq}\\ 
     \colorbox{blue!15}{block:block\_touch\_buffer}\\
     \colorbox{blue!15}{block:block\_unplug}}&
     \makecell[l]{cachefiles:cachefiles\_create\\   
     cachefiles:cachefiles\_lookup\\
     cachefiles:cachefiles\_mark\_active\\
     \colorbox{blue!15}{filemap:mm\_filemap\_add\_to\_page\_cache}\\} \\
     
         \hline
    Scheduler & \makecell[l]{\colorbox{blue!15}{sched:sched\_process\_exec}\\  
     \colorbox{blue!15}{sched:sched\_process\_free}\\
     \colorbox{blue!15}{sched:sched\_process\_wait}\\
     sched:sched\_switch\\
     }&
     \makecell[l]{
     signal:signal\_deliver\\
     \colorbox{blue!15}{signal:signal\_generate}\\
     \colorbox{blue!15}{task:task\_newtask}\\
     \colorbox{blue!15}{cpu-migrations}\\}&
     \makecell[l]{
     cs\\
     alarmtimer:alarmtimer\_fired\\
     alarmtimer:alarmtimer\_start}\\

    \hline
    CPU & \makecell[l]{
     \colorbox{blue!15}{clk:clk\_set\_rate}\\
     rpm:rpm\_resume\\}&
     \makecell[l]{
     rpm:rpm\_suspend\\}&
     \makecell[l]{    
     \colorbox{blue!15}{ipi:ipi\_raise}} \\   
    
    \hline
    Device Drivers & \makecell[l]{\colorbox{blue!15}{irq:irq\_handler\_entry}\\
     mmc:mmc\_request\_start\\}&
     \makecell[l]{
     preemptirq:irq\_enable\\
     gpio:gpio\_value\\}&
     \makecell[l]{
     dma\_fence:dma\_fence\_init }\\
    \hline
    
    Random Numbers & \makecell[l]{\colorbox{blue!15}{random:get\_random\_bytes}\\} &\makecell[l]{\colorbox{blue!15}{random:mix\_pool\_bytes\_nolock}\\} & \makecell[l]{\colorbox{blue!15}{ random:urandom\_read}} \\

    \hline
    \end{tabular}
    \label{tab:features}
\end{table*}

The online process is periodically executed to detect cyberattacks affecting the behavior of resource-constrained spectrum sensors. With that goal in mind, when the \textit{Model Evaluation} component has the models it evaluates the received periodic behavior vectors (step 4 in \figurename~\ref{fig:architecture}). Each evaluation predicts if the current behavior vector is abnormal, which means that the spectrum sensor is infected.

%--------------------------------------------------------

\section{Validation Scenario \& Experiments}
\label{sec:validation}

ElectroSense~\cite{rajendran:2018:electrosense} is the crowdsensing RF spectrum monitoring platform selected to validate the CyberSpec framework and to measure its performance, when detecting anomalies produced by the SSDF attacks of Section~\ref{sec:attacks}. 

%\subsection{ElectroSense as Validation Scenario}

%ElectroSense is a real-life IoT-oriented crowdsensing platform in charge of sensing the RF spectrum to provide different services, such as spectrum visualization and decoding. ElectroSense has deployed globally resource-constrained, inexpensive, and easily accessible sensors, such as Raspberry Pi equipped with software-defined radio (SDR) kits to sense the spectrum. Each sensor is connected to the internet and periodically sends RF spectrum data to a backend platform. Apart from receiving, processing, analyzing, and visualizing spectrum data, the ElectroSense backend can also control some functionalities of the software running on top of the Raspberry Pis (ElectroSense sensors). At this point, it is worth mentioning that the modules of software executed on the ElectroSense sensors are open source and publicly available.

As validation scenario, this work has deployed a set of Raspberry Pis equipped with SDR kits and the ElectroSense software sensing the RF spectrum. Sensors monitor the PSD values of frequency bands segments ranging from 20 MHz to 1.6 GHz in blocks of 2.4MHz. This process takes about 50 s. The following Raspberry Pi acting as ElectroSense sensors have been considered:

\begin{itemize}
    \item \textit{Six Raspberry Pis 3 Model B} with Quad Core 1.2 GHz Broadcom BCM2837 64-bit CPU, and 1 GB of RAM.
    \item \textit{Three Raspberry Pis 4} with Quad Core 1.5 GHz Broadcom BCM2711 64-bit CPU, and 2 GB of SDRAM.
\end{itemize}

These nine spectrum sensors are connected to the Internet through the next four local area networks, which are deployed in different geographical locations.

\begin{itemize}
    \item \textit{LAN\_1}: One Raspberry Pi 3.
    \item \textit{LAN\_2}: One Raspberry Pi 4.
    \item \textit{LAN\_3}: Two Raspberry Pis 3 \& one Raspberry Pi 4.
    \item \textit{LAN\_4}: Three Raspberry Pis 3 \& one Raspberry Pi 4.
\end{itemize}

As indicated in Section~\ref{sec:framework}, each sensor implements the Behavior Fingerprinting module of CyberSpec to monitor and send its behavior. The computational and time costs of the module are shown in \tablename~\ref{tab:consumption}. In particular, the module monitors the sensor behavior in time windows of 50 s and requires 6.8 s to pre-process the events. 50 s is the minimum monitoring time required to capture all SSDF attacks affecting any RF segment band since it is the time needed by the sensors to scan the spectrum. In addition and for simplicity, the Cyberattack Detection module, in charge of creating the datasets, training and evaluating ML/DL models, is deployed on a trusted server.

\begin{table}[ht]
\centering
\begin{center}
\caption{Behavior Fingerprinting Module Cost for Raspberry Pis 3 and 4}
\begin{tabular}{c|c|c|c|c}
\textbf{CPU} & \textbf{Memory} & \textbf{Storage} & \textbf{Monitoring} & \textbf{Processing}  \\ \hline
\makecell[l]{0.5-2\% 1 Core \\ with 8\% peaks} & 900 kB & 7.8 kB & 50 s & 6.8 s \\ \hline
\end{tabular}
\label{tab:consumption}
\end{center}
\end{table}

In such a scenario, the normal behavior of each Raspberry Pi was monitored for eight days, and a dataset per device was created. After that, several configurations of the seven attacks defined in Section~\ref{sec:attacks} were sequentially executed in each Raspberry Pi for two hours, creating one dataset per attack configuration and device. Attacks configurations differ in the affected spectrum bandwidth to simulate and hide heterogeneous wireless communications standards. In particular, the selected configurations of attacks affect transmissions technologies such as WiFi (from 20 MHz to 160 MHz depending on the 802.11 version), Bluetooth (2 MHz), 3G (200 kHz), and SOS (20 kHz). At this point, it is worth mentioning that each Raspberry Pi has been infected with only one attack at the same time. As a summary, the following datasets were created per sensor.

\begin{itemize}
    \item One dataset with normal behavior (192 h of monitoring).
    \item Forty-two datasets with malicious behavior (2h of monitoring each one). In detail, six datasets with a different bandwidth configuration (20 kHz, 200 kHz, 2 MHz, 20 MHz, 80 MHz, and 160 MHz) per attack (Noise, Spoof, Repeat, Confusion, Mimic, Freeze, and Delay). 
\end{itemize}

\begin{table*}[ht]
	\caption{Anomaly Detection Algorithms and Hyper-parameters Tested \& Selected}
	\centering
    \begin{tabular}{m{2cm}|m{6cm}|m{4cm}}
        \textbf{Algorithm} & \textbf{Hyperparameters tested} & \textbf{Hyperparameters selected}\\
        \hline
        Autoencoder & $layers\in [1,3], neurons\_layer\in [10,60]$ & $layers=1, neurons=40$\\
        \hline
        LOF & $n\_neighbors\in [3,25]$ & $n\_neighbors=15$\\
        \hline
        OC-SVM & \makecell[l]{$gamma\in [0.001,100]$\\$kernel\in \{'rbf', 'linear', 'sigmoid','poly'\}$\\$degree\in [2,5](only\ poly\ kernel)$} & $kernel='rbf',gamma=0.001$\\
        \hline
        IF & $Number\_of\_trees \in [50,1000]$ & $Number\_of\_trees=150$\\
        \hline
        COPOD & - & -\\
        \hline
    \end{tabular}
    \label{tab:anomaly_alg_hyp}
\end{table*}

\subsection{Experiments}

This section proposes the next three experiments to evaluate the performance of CyberSpec when detecting the previous configurations of SSDF attacks (see Section \ref{sec:attacks}).

\begin{itemize}
    \item \textbf{Exp\_1}: Performance analysis of individual models per Raspberry Pi.
    \item \textbf{Exp\_2}: Performance analysis of models per type of device, with different amounts of Raspberry Pis 3 used during training and evaluation.
    \item \textbf{Exp\_3}: Performance analysis of global models combining the behavior of all Raspberry Pis 3 and 4.
\end{itemize}

To perform the three experiments, a common methodology has been followed. It starts choosing the datasets modeling the normal behavior of the selected Raspberry Pis (different for each experiment). After that, the pipeline indicated in Section~\ref{sec:framework} is followed to prepare the data for the ML/DL phase. Next, a set of unsupervised anomaly detection algorithms are selected. In particular, Autoencoder, Isolation Forest (IF), Copula-Based Outlier Detection (COPOD), Local Outlier Factor (LOF), and One-Class Support Vector Machine (OC-SVM) with a diverse set of hyper-parameters are considered at this stage (see \tablename~\ref{tab:anomaly_alg_hyp}). The next step is to adjust the threshold to detect anomalies by using a well-known statistical approach. In this sense, the Interquartile Rule \cite{vinutha:2018:iqr} is applied to find outliers. This technique calculates the quartiles and the interquartile range (IQR) in the scores of the training instances and defines two thresholds calculated as \textit{Q1 - 1.5*IQR} and \textit{Q3 + 1.5*IQR}. Then, any score below or over these values is treated as an anomaly. In the case of the Autoencoder, the Mean Squared Error (MSE) of the input reconstruction (difference between model output and input in each feature) was used as a score, and in the case of LOF, IF, COPOD, and OC-SVM, the raw algorithm outputs were used. After that, a final selection of the hyper-parameters and the threshold is made using the validation dataset (see \tablename~\ref{tab:anomaly_alg_hyp}). At this point, it is important to mention that the hyper-parameters selection for the three experiments has been made using the configuration of Exp\_3, due to the minimum impact of having individuals per experiment and the number of models (more than 100 trained across all experiments). Finally, the detection performance of each algorithm is evaluated with the normal and the behavior under-attack of each device (in each experiment, different data setups are used for training and testing). The metrics used to compare the detection performance are the TNR (True Negative Rate) and TPR (True Positive Rate). TNR indicates the number of non-anomalies found in the normal or benign behavior, while TPR is used to evaluate malicious behavior and provides the number of detected anomalies.

\subsubsection{Exp\_1: Individual models}

This experiment employs the nine Raspberry Pis (six Raspberry Pis 3 and three Raspberry Pis 4) and their training datasets (modeling the 72\% of their normal behaviors) to train the five selected algorithms (Autoencoder, IF, LOF, COPOD, and OC-SVM) per device (45 ML/DL models in total). After that, the individual models of each device have been evaluated with \textit{(i)} the testing dataset (modeling the 10\% of normal behavior) of that device, and \textit{(ii)} the 42 datasets modeling the different configurations of the SSDF attacks affecting that device.

\begin{figure}
    \centering
    \includegraphics[width=\columnwidth]{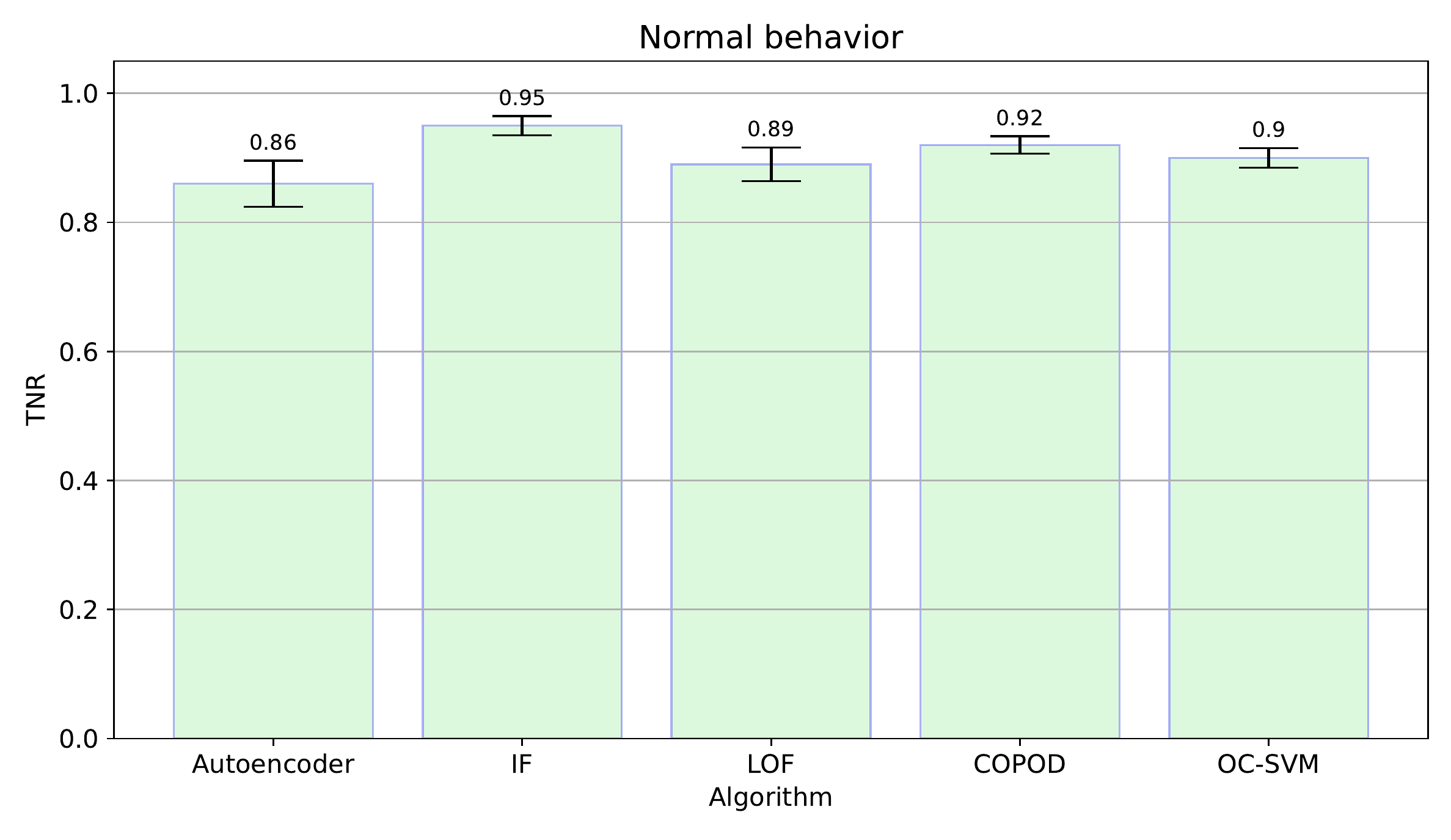}
    \caption{Average TNR Performance of Individual ML/DL Models When Detecting Normal Behavior}
    \label{fig:exp1_normal}
\end{figure}

For each algorithm, \figurename~\ref{fig:exp1_normal} shows the TNR mean and standard deviation of each individual model algorithm when they are evaluated with normal behavior. As can be seen, IF is the model obtaining the best average performance for all devices (95\% TNR). At this point, it is important to mention that since the evaluated datasets do not contain anomalies generated by SSDF attacks, the TPR metric is not considered when detecting normal behavior. 

In addition, to choose the algorithm providing the best performance is needed to analyze the performance when detecting SSDF attacks. In this sense, \figurename~\ref{fig:exp1_attack} shows the mean TPR (for the nine devices) of the five individual ML/DL models.

\begin{figure}[ht]
    \centering
    \includegraphics[width=\columnwidth]{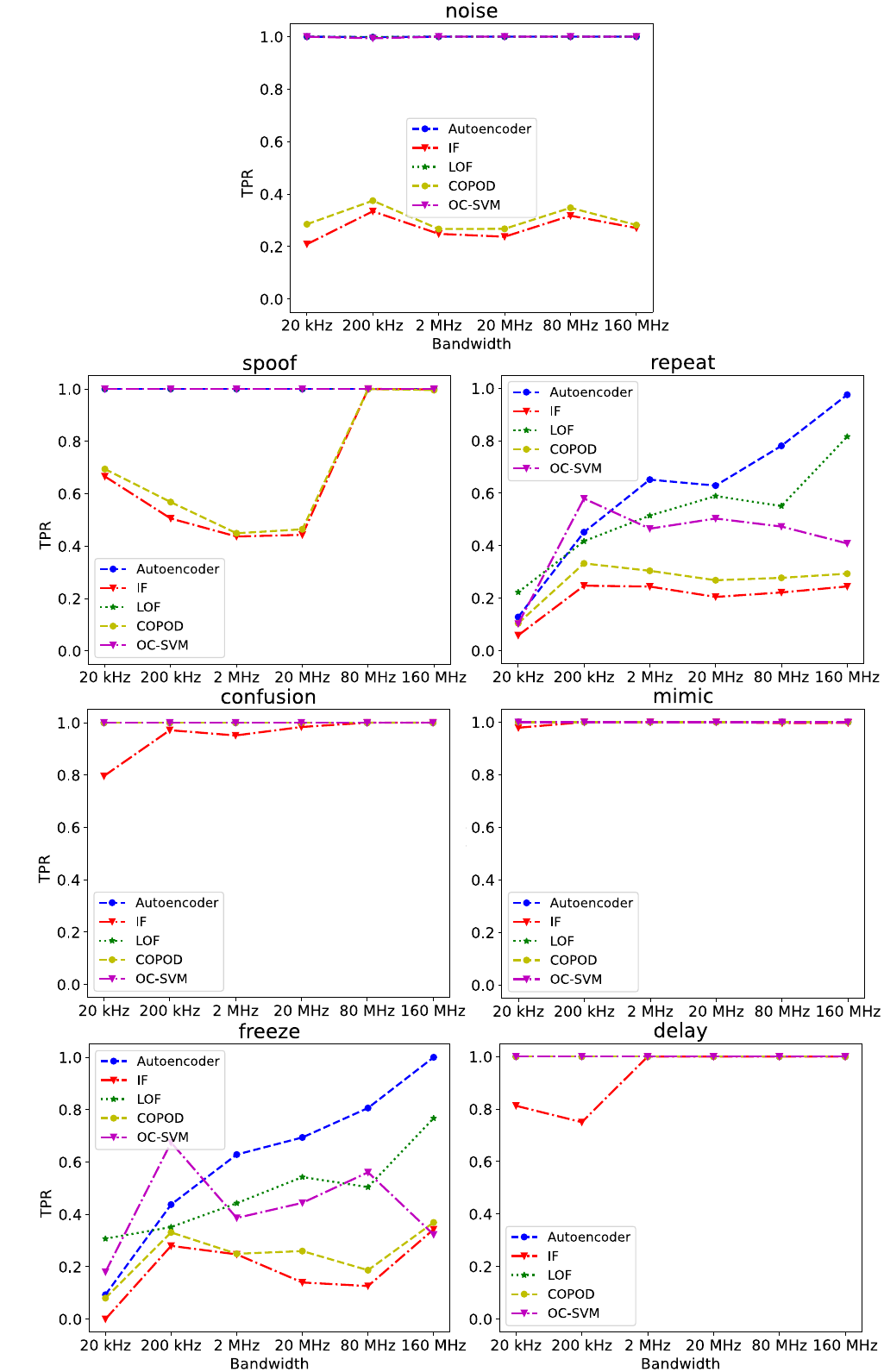}
    \caption{Average TPR Performance of Individual ML/DL Models When Detecting SSDF Attacks}
    \label{fig:exp2_attacks}
\end{figure}

As can be seen in \figurename~\ref{fig:exp1_attack}, in general, Autoencoder, LOF, and OC-SVM are the models providing the best performance (100\% TPR) for all configurations of noise, spoof, confusion, mimic and delay). Repeat and Delay attacks deserve special consideration since they are not adequately detected until the affected bandwidth reaches 80 and 160 MHz (only by Autoencoder). It is due to the selected features do not monitor events produced by file read operations (internal behavior of these two attacks, as shown in Section~\ref{sec:attacks}). In this sense, when the affected bandwidth is sufficient (80 and 160 MHz), the attack impacts the rest of the monitored events, and these variations are slightly detected.

\subsubsection{Exp\_2: Device-type models}

Since individual models per device might be cumbersome to maintain in dynamic platforms, such as ElectroSense with a significant number of new devices per month, this experiment evaluates the performance of models per device family or type (one model for Raspberry Pis 3 and another for Raspberry Pis 4). In particular, it analyzes the fact of having device-type models \textit{(i)} trained and evaluated with the normal behavior of all devices, and \textit{(ii)} trained with different numbers of devices and evaluated with the rest.

\begin{figure}[htpb!]
    \centering
    \includegraphics[width=\columnwidth]{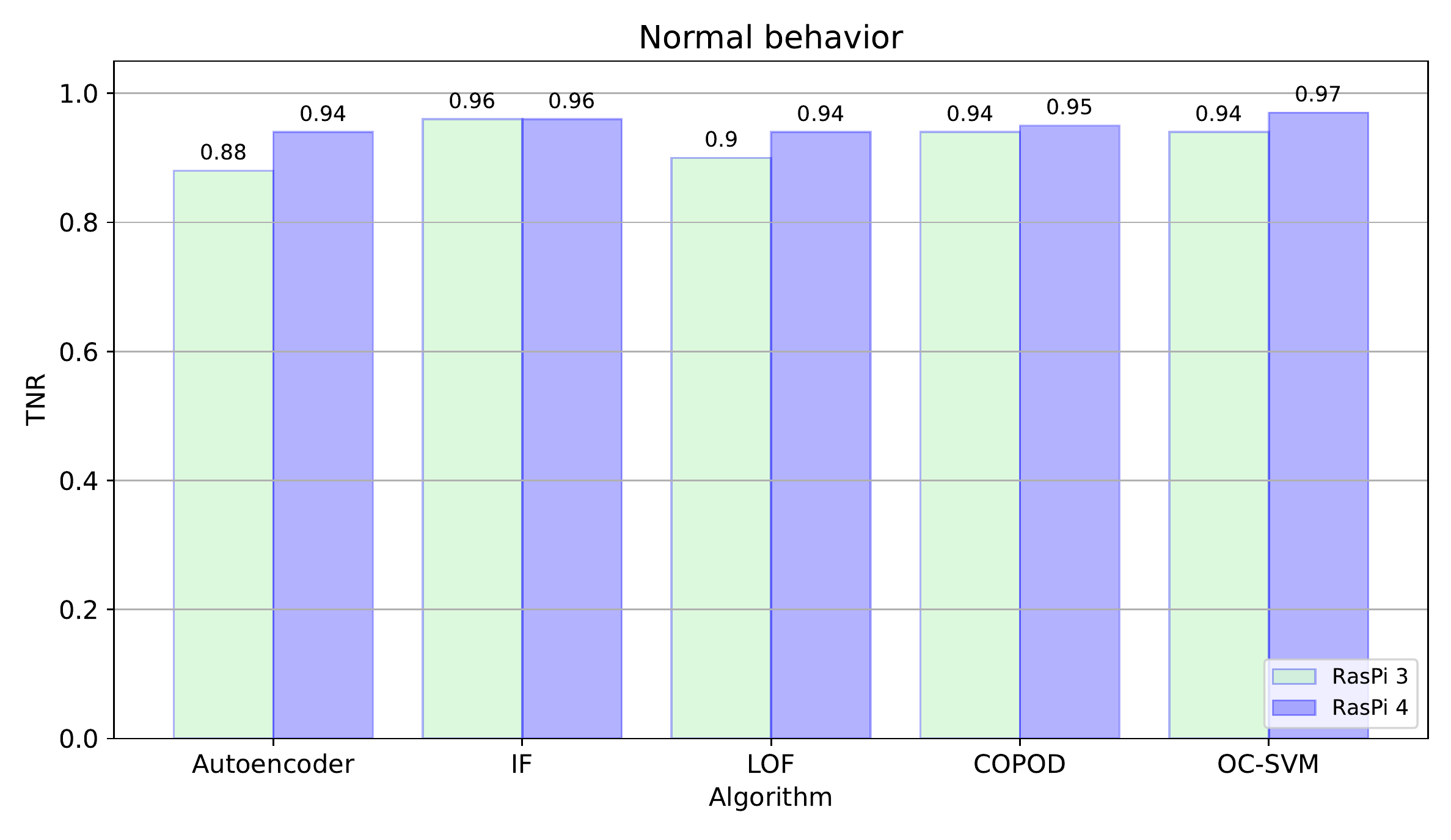}
    \caption{TNR Performance of Device-type ML/DL Models for Raspberry Pi 3 and 4 Recognizing Normal Behavior}
    \label{fig:exp2_normal}
\end{figure}

Regarding the first setup, the training datasets (modeling the 72\% of the normal behavior) of the 100\% of Raspberry Pis 3 (six devices) are combined to feed the five selected algorithms per device family. After that, the testing datasets of the six Raspberry Pis 3 (modeling the 10\% of their normal behavior) are concatenated and evaluated. This process is repeated for the three Raspberry Pis 4. \figurename~\ref{fig:exp2_normal} shows the TNR for each algorithm and device type (light green for Raspberry Pis 3 and blue for Raspberry Pis 4). As can be seen, for both families of devices, the performance is similar (being Autoencoder the one obtaining greater differences for both families with 0.85\% and 0.93\% TNR), showing that the proposed features and framework are suitable for Raspberry Pis 3 and 4. In addition, comparing these results with the obtained by the individual models of the previous experiment, the differences are minimal as well.

\begin{figure*}
     \centering
     \subfloat[][Raspberry Pi 3]{\includegraphics[width=0.5\textwidth]{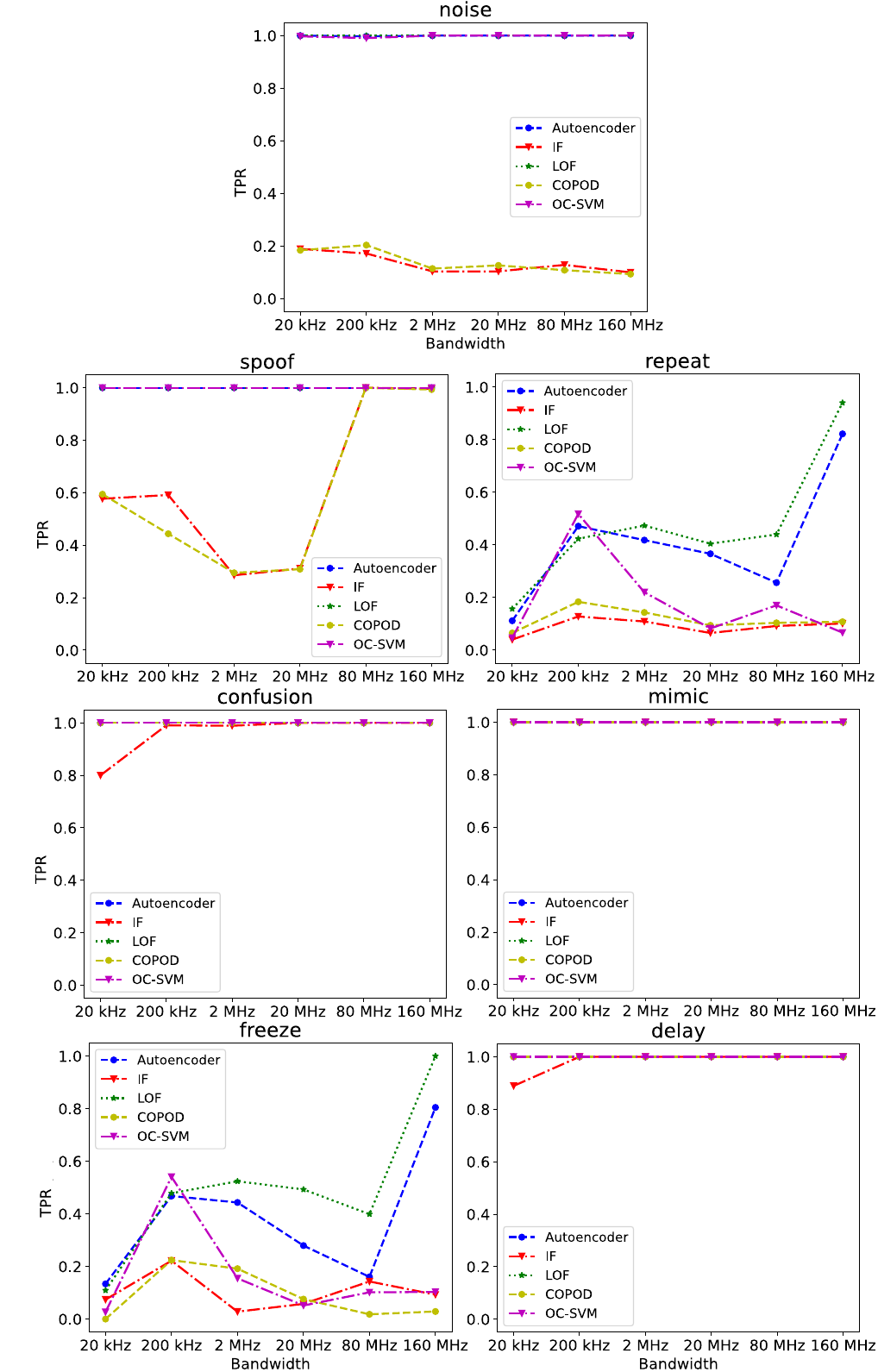}}
     \subfloat[][Raspberry Pi 4]{\includegraphics[width=0.5\textwidth]{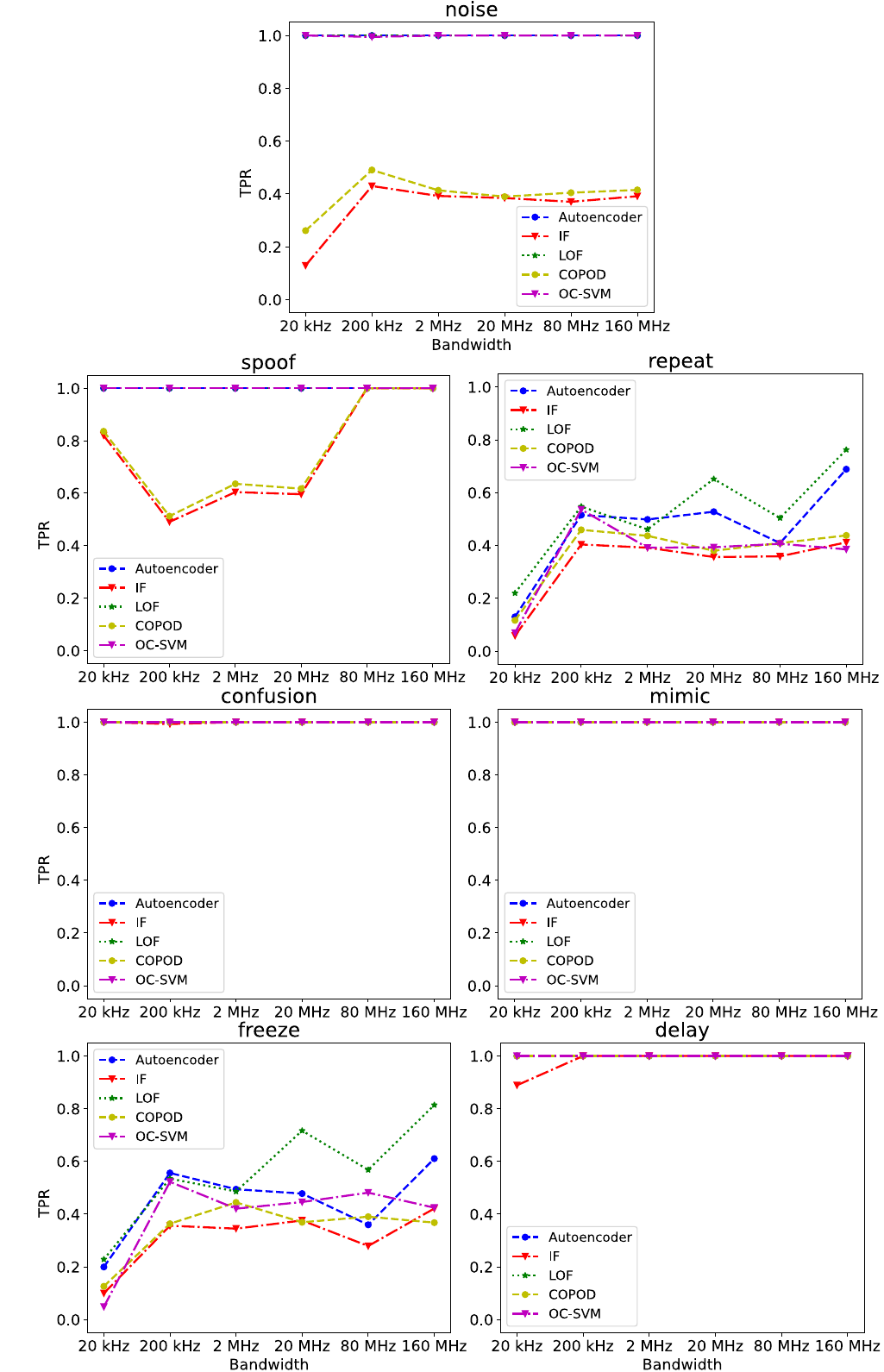}}
     
     \caption{Average TPR Performance of device-type ML/DL Models When Detecting SSDF Attacks}
     \label{fig:exp1_attack}
\end{figure*}

Regarding the detection of attacks, \figurename~\ref{fig:exp2_attacks} shows the mean TPR of the Raspberry Pi 3 \& 4 family models when detecting anomalies contained in the 42 malicious datasets per device (252 datasets of Raspberry Pis 3, and 126 of Raspberry Pis 4). These results are similar to those shown in \figurename~\ref{fig:exp1_attack}. In particular, all configurations of noise, spoof, confusion, mimic, and delay are detected by Autoencoder, LOF, and OC-SVM. These results show a promising path to have models per family of devices, reducing the number of models in the platform and, therefore, the maintenance cost.

\begin{figure}[htpb!]
    \centering
    \includegraphics[width=\columnwidth]{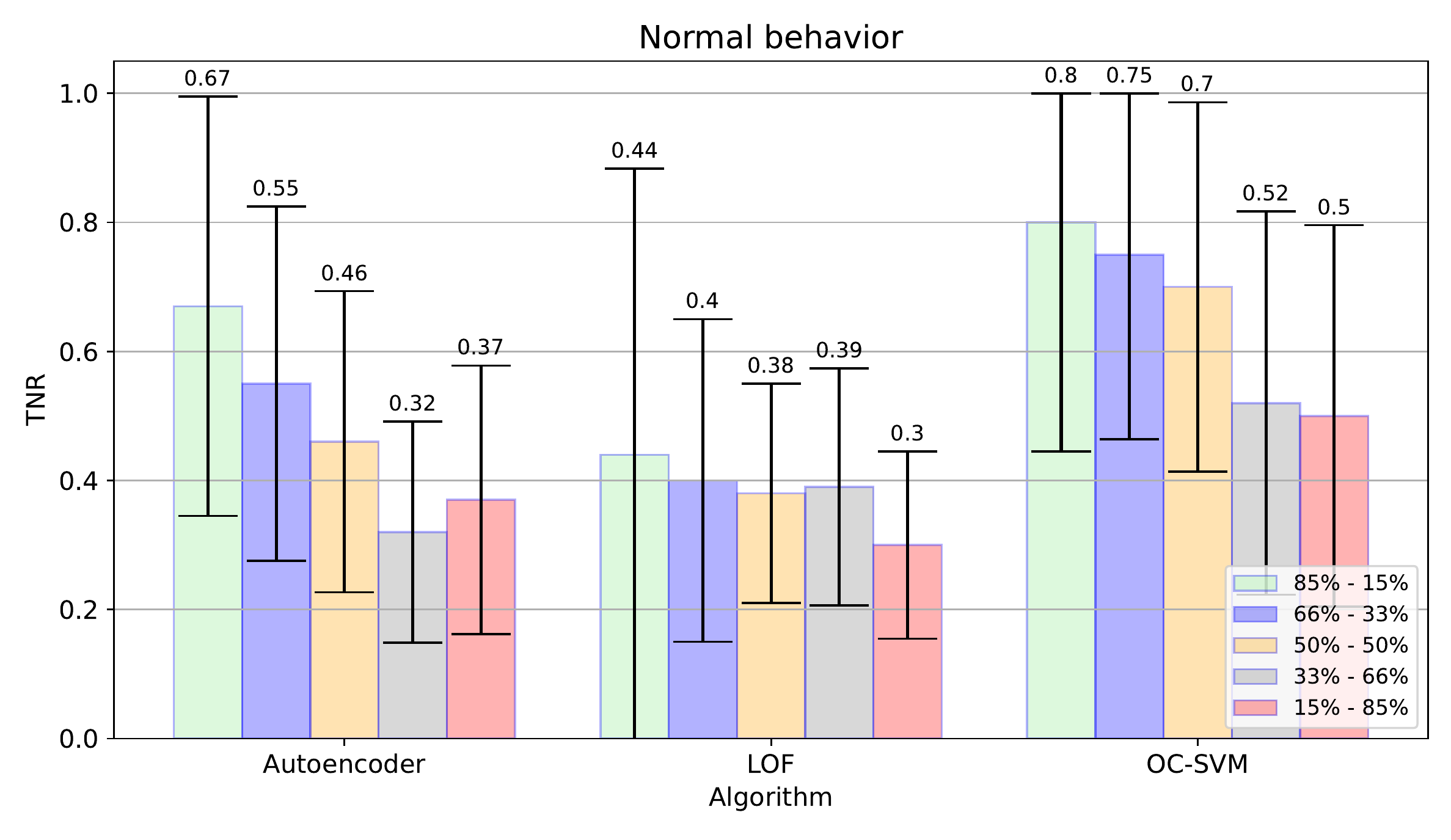}
    \caption{Average TNR Performance of Device-type ML/DL Models Trained with Different Numbers of Raspberry Pis 3 When Detecting Normal Behavior}
    \label{fig:exp2_1_normal}
\end{figure}

Once the suitability of device type models is demonstrated, the second setup of this experiment evaluates the scalability of the proposed solution when new devices appear in the platform (unseen during the training phase). In this sense, different tests have been performed excluding the 15\%, 33\%, 50\%, 66\%, and 85\% of Raspberry Pis 3 (1, 2, 3, 4, and 5 devices, respectively) from training. For each test, all combinations of devices have been performed to show robust results. In other words, and as an example, when one device is excluded from the training phase, all combinations (6 in total) have been considered to train the models. At this point, it is important to mention that only Raspberry Pis 3 have been selected due to the number of available devices. As in the previous experiment, 72\% of the normal behavior of the selected devices has been used for training and 100\% of the normal behavior of excluded devices for evaluation. \figurename~\ref{fig:exp2_1_normal} shows the TNR mean and standard deviation of Autoencoder, LOF, and OC-SVM. Only these three algorithms are shown in the figure do to they are the ones obtaining 100\% TPR for all configurations of noise, spoof, confusion, mimic, and delay (in this experiment test and the previous ones).

As can be seen in \figurename~\ref{fig:exp2_1_normal}, regardless of the number of excluded devices, OC-SVM is the model providing the best averaged TNR. In addition, and as expected, the higher the number of devices excluded from the training, the lower the TNR. This is due to the fact of having independent data with different distributions (also known as non-IID) per device. These different data distributions could be influenced by external factors such as network bandwidth, traffic, device temperature, or processes. In any case, OC-SVM obtains good TNR averages (80\%, 75\%, and 70\%) when 85\%, 66\%, and 50\% of devices are used only during training. \figurename~\ref{fig:exp2_1_normal} also shows that the OC-SVM models work very well for the majority of devices (high TNR mean) but bad for a few of them (high standard deviation). Furthermore, when more than 50\% of devices are excluded from training, the detection performance of OC-SVM and all models drops to 50\%. In conclusion, the scalability results obtained by OC-SVM are promising, but more Raspberry Pis 3 connected to the same and different networks are required to confirm the positive trend.

\subsubsection{Exp\_3: Global model}

This last experiment consists of training a global model that combines the normal behavior of all Raspberry Pis 3 and 4. For that, it combines the nine datasets modeling the 72\% of the devices normal behavior, and the samples of each dataset were balanced for the training process of the five selected algorithms. Finally, the resulting models were evaluated with both normal (testing dataset of each device with 10\% of samples) and under attack behaviors of all devices. 

\begin{figure}[htpb!]
    \centering
    \includegraphics[width=\columnwidth]{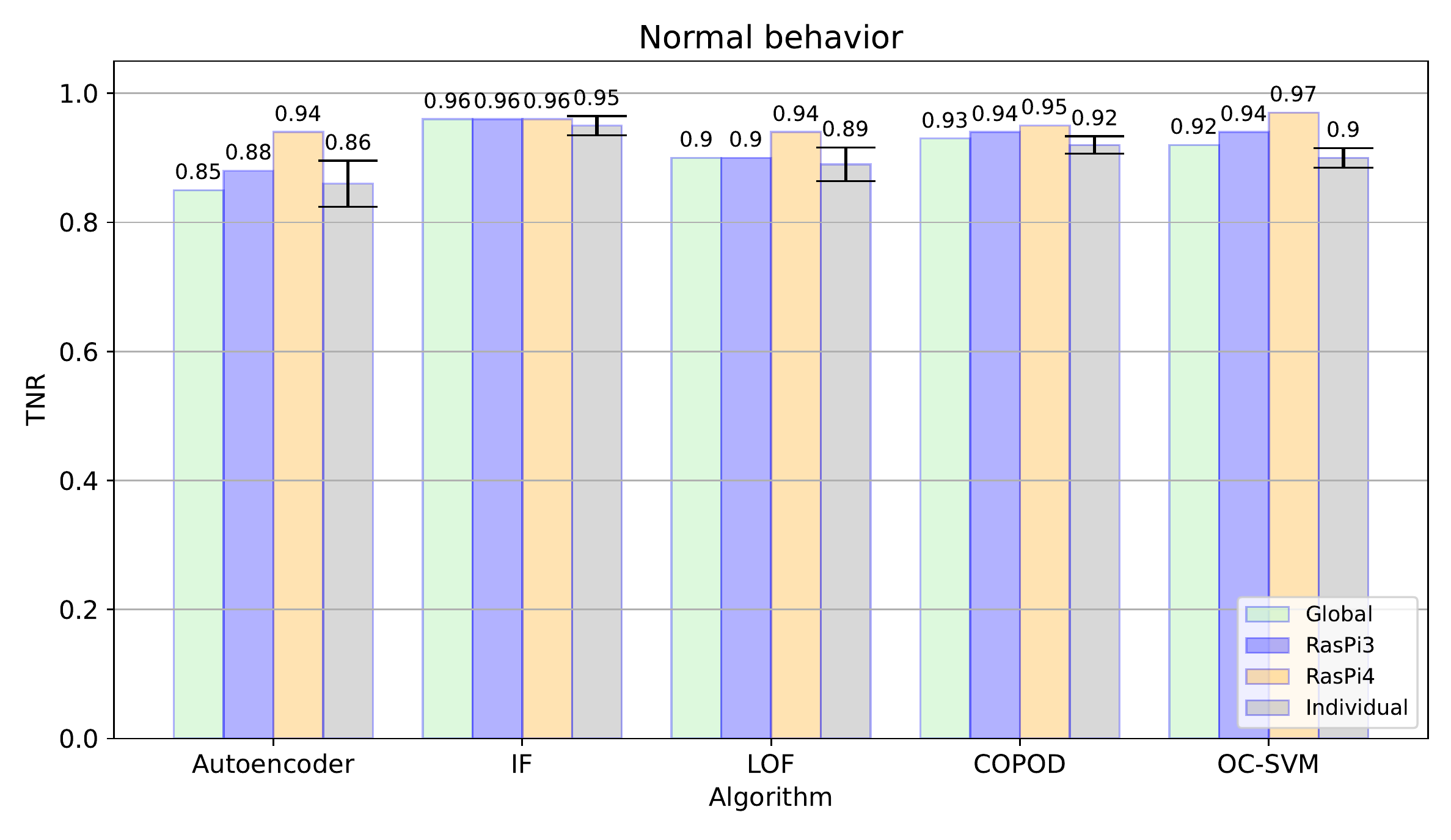}
    \caption{Comparative of the TPR Performance of Different ML/DL Models When Detecting Normal Behavior}
    \label{fig:exp3_normal}
\end{figure}

\figurename~\ref{fig:exp3_normal} shows the TNR of global models (light green bars), models-type for Raspberry Pi 3 (blue bars) and Raspberry Pis 4 (orange bars), and individual models for a Raspberry Pi 4 (gray bars) when detecting normal behaviors. As shown, there are no great differences in the TNR of the three experiments, indicating that models combining the normal behavior of Raspberry Pis 3 and 4 perform reasonably well when detecting normal behaviors. Regarding the detection of anomalies, the performance of the five ML/DL algorithms was evaluated. Still, for the sake of simplicity, \tablename~\ref{tab:datasets} shows the TPR obtained by OC-SVM, the one providing the best ratio of TNR/TPR performance. As it can be seen, spoof, confusion, mimic, and delay are perfectly detected in all their configurations. The noise attack was perfectly detected almost for all configurations, excluding the less impactful one (20kHz). Finally, both Repeat and Freeze were not adequately detected, as in the previous two experiments. It is important to mention that these results have been calculated by computing the average mean of the TPR score provided by each Raspberry Pi 3 and 4.

\begin{table}[ht]
\centering
\begin{center}
\caption{Average TPR of OC-SVM Global Model When Detecting SSFD Attacks}
\begin{tabular}{l|c|c|c|c|c}
\textbf{Attack} & \textbf{20 kHz} & \textbf{200 kHz} & \textbf{2 MHz} & \textbf{80 MHz} & \textbf{160 MHz} \\ \hline
 noise & 61\% & 99\% & 100\% & 100\% & 100\% \\ \hline
 spoof & 99\% & 100\% & 100\% & 100\% & 100\%\\ \hline
 repeat & 6\% & 29\% & 26\% & 18\% & 26\%\\ \hline
 confusion & 100\% & 100\% & 100\% & 100\% & 100\%\\ \hline
 mimic & 100\% & 100\% & 100\% & 100\% & 100\%\\ \hline
 freeze & 7\% & 24\% & 30\% & 28\% & 32\% \\ \hline
 delay & 100\% & 100\% & 100\% & 100\% & 100\%\\ \hline
\end{tabular}
\label{tab:datasets}
\end{center}
\end{table}

Finally, \tablename~\ref{tab:time} shows the time needed by the server (equipped with an Intel i7-5930K CPU @ 3.50GHz, 3 NVIDIA GTX1080 GPUs and 96 GB RAM) to train and evaluate each model for the global experiment (worst-case scenario because the individual and device-type experiments need less data for training). The testing time is the most important because it affects the detection time of CyberSpec, which is the sum of the monitoring time (50 s), the time to pre-process the monitored events and create the features (6.8 s), and the time to evaluate the features (2.75 s worst-case scenario). In summary, CyberSpec takes less than 60 s to detect attacks since the model training process can be done offline, and the communication time between the sensors and the server where models are evaluated is minimal.

\begin{table}[ht]
\centering
\begin{center}
\caption{Time Needed to Train and Test Global Models}
\begin{tabular}{l|c|c|c|c|c}
& Autoencoder & IF & LOF & COPOD  & OC-SVM  \\ \hline
\textbf{Train} & 97.82 s & 10.99 s & 58.27 s & 6.08 s & 770.39 s \\ \hline
\textbf{Test} & 0.03 s & 0.04 s & 0.02 s & 0.02 s & 0.01 s \\ \hline
\end{tabular}
\label{tab:time}
\end{center}
\end{table}

%--------------------------------------------------------

\section{Summary, Conclusions, and Future Work}
\label{sec:conclusions}

This work introduced and formally described the behavior of seven Spectrum Sensing Data Falsification (SSDF) attacks (Repeat, Mimic, Confusion, Noise, Spoof, Freeze, and Delay) affecting crowdsensing spectrum sensors. The second contribution was the design and implementation of CyberSpec, an ML/DL-oriented framework that monitors internal events of IoT devices, such as network, virtual memory, files system, scheduler, system calls, CPU, device drivers, and random numbers, to detect anomalies produced by cyberattacks. The performance of CyberSpec has been evaluated in a real crowdsensing spectrum monitoring platform called ElectroSense, where nine Raspberry Pis 3 and 4 acting as spectrum sensors were infected with the previous SSDF attacks. Three experiments with different unsupervised ML/DL models were performed to evaluate the CyberSpec detection capabilities. Individual ML/DL models per device, models per family of devices (excluding different \% of devices from training), and global models combining all devices have provided a promising performance when detecting the normal behavior of six Raspberry Pis 3 and three Raspberry Pis 4. Five (Noise, Spoof, Confusion, Mimic, and Delay) of the seven analyzed attacks are almost perfectly detected (100\% TPR) by Autoencoder, LOF, and OC-SVM in the three experiments. The fact of having models combining normal behaviors of different devices does not reduce the detection performance significantly. Moreover, looking at the scalability of device-type models, when the 15\%, 33\%, and 50\% of devices are excluded from training, they perform relatively well for most excluded devices (80-70\% TPR and 100\% TNR for five attacks). Repeat and Freeze are not correctly detected because CyberSpec does not monitor read file operations, and the behavior of these attacks relies on that. Finally, the CyberSpec framework needs less than 60 s to detect normal/under-attack behavior while keeping a reduced computational cost in the Raspberry Pis.

With the goal of giving an answer to the open challenges depicted in Section \ref{sec:intro}, the main conclusions after performing the experiments and analyzing the results are \textit{(i)} the proposed features and individual ML/DL models per device are appropriate to detect precisely both normal behaviors and anomalies produced by attacks affecting IoT devices; \textit{(ii)} global ML/DL models (per family of devices and combining two families) show a promising detection performance in scenarios with a reduced number of ElectroSense sensors, but more experiments are needed to evaluate its scalability when the number of devices increases; and \textit{(iii)} additional behavioral events and features are required to properly detect attacks, such as Repeat and Freeze, having a slight impact on the internal behavior of the ElectroSense sensors.

Future work will evaluate the performance of privacy-preserving mechanisms. Solutions like the proposed in this work, where global models are trained with the behavior of different devices, are not suitable for privacy-preserving scenarios where federated learning can be a solution to train models while preserving data sensitiveness. Furthermore, supervised algorithms will be investigated to classify cyberattacks in different families. Finally, adding new event families will lead to the improvement of the detection accuracy, especially in attacks such as Repeat and Freeze.

\section*{Acknowledgment}

This work has been partially supported by \textit{(a)} the Swiss Federal Office for Defense Procurement (armasuisse) with the CyberSpec (CYD-C-2020003) and TREASURE (R-3210/047-31) projects and \textit{(b)} the University of Zürich UZH.

\bibliographystyle{unsrt}  
\bibliography{references}

\begin{IEEEbiography}[{\includegraphics[width=1in,clip]{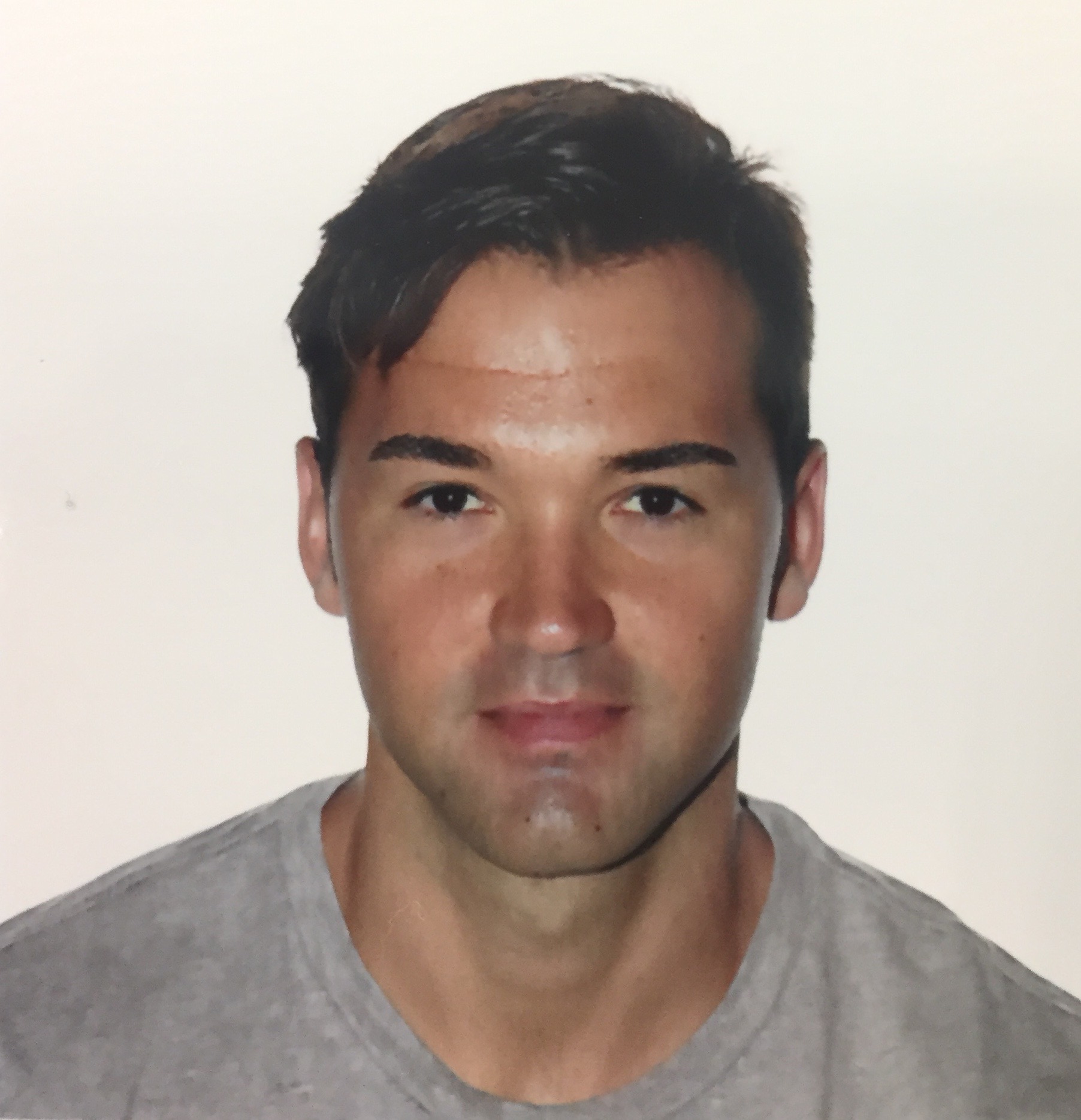}}]{Alberto Huertas Celdrán} received the MSc and PhD degrees in Computer Science from the University of Murcia, Spain. He is currently a postdoctoral fellow at the Communication Systems Group CSG, Department of Informatics IfI at the University of Zurich UZH. His scientific interests include medical cyber-physical systems (MCPS), brain-computer interfaces (BCI), cybersecurity, data privacy, continuous authentication, semantic technology, context-aware systems, and computer networks.
\end{IEEEbiography}

\begin{IEEEbiography}[{\includegraphics[width=1in,clip]{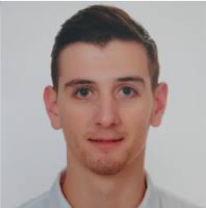}}]{Pedro M. Sánchez Sánchez} received the MSc degree in computer science from the University of Murcia, Spain. He is currently pursuing his PhD in computer science at University of Murcia. His research interests are focused on continuous authentication, networks, 5G, cybersecurity and the application of machine learning and deep learning to the previous fields.
\end{IEEEbiography}

\begin{IEEEbiography}[{\includegraphics[width=1in,clip]{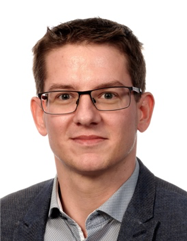}}]{Gérôme Bovet} is the head of data science for the Swiss Department of Defense, where he leads a research team and a portfolio of about 30 projects. His work focuses on Machine and Deep Learning approaches applied to cyber-defense use cases, with an emphasis on anomaly detection, adversarial and collaborative learning. He received his PhD in networks and systems from Telecom ParisTech, France, in 2015.
\end{IEEEbiography}

\begin{IEEEbiography}[{\includegraphics[width=1in,clip]{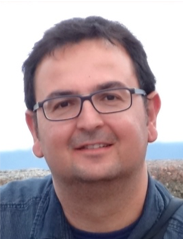}}]{Gregorio Martinez Pérez} is Full Professor in the Department of Information and Communications Engineering of the University of Murcia, Spain. His scientific activity is mainly devoted to cybersecurity and networking, also working on the design and autonomic monitoring of real-time and critical applications and systems. He is working on different national (14 in the last decade) and European IST research projects (11 in the last decade) related to these topics, being Principal Investigator in most of them. He has published 160+ papers in national and international conference proceedings, magazines and journals.
\end{IEEEbiography}

\begin{IEEEbiography}[{\includegraphics[width=1in,clip]{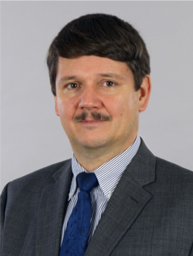}}]{Burkhard Stiller}  received the Informatik-Diplom (MSc) degree in Computer Science and the Dr. rer.- nat. (PhD) degree from the University of Karlsruhe, Germany, in 1990 and 1994, respectively. He was with the Computer Lab, University of Cambridge, U.K, ETH Zürich, Switzerland, and the University of Federal Armed Forces Munich, Germany. Since 2004 he chairs the Communication Systems Group CSG, Department of Informatics IfI, University of Zürich UZH, Switzerland as a Full Professor. Besides being a member of the editorial board of the IEEE Transactions on Network and Service Management, Springer’s Journal of Network and Systems Management, and the KICS’ Journal of Communications and Networks, Burkhard is past Editor in-Chief of Elsevier’s Computer Networks journal. His main research interests are published in well over 300 research papers and include systems with a fully decentralized control (blockchains, clouds, peer-to-peer), network and service management (economic management), Internet-of-Things (security of constrained devices, LoRa), and telecommunication economics (charging and accounting).
\end{IEEEbiography}

\end{document}